\documentclass{aa}  

\usepackage{graphicx}
\usepackage{booktabs}
\usepackage[utf8]{inputenc}

\usepackage{txfonts}
\usepackage{xcolor}

\newcommand{\teff}{$T_{\rm eff}$}
\newcommand{\logg}{$\log\,{g}$}
\newcommand{\vsini}{$v\sin{i}$}

\usepackage{hyperref}

\begin{document}

   \title{Detectability of deuterium in spectra of early-type stars}

\author{
  V. Mitrokhina\inst{1} \and
  C. P. Folsom\inst{1} \and
  M. Kama\inst{2,1} \and
  A. Aret\inst{1} 
}

\institute{
  Tartu Observatory, University of Tartu, Observatooriumi 1, Tõravere, 61602 Tartumaa, Estonia \\
  \email{veronika.mitrokhina@ut.ee}
  \and
  Department of Physics and Astronomy, University College London, Gower Street, London, WC1E 6BT, UK \\
}

   \date{Received  ; accepted  }

  \abstract  
   {Deuterium is easily destroyed in stellar interiors through nuclear fusion. It is therefore usually not expected to be present in stellar photospheres. Early-type stars, with radiative envelopes that mix slowly, may provide a favourable environment for the survival of recently accreted deuterium. }
   {In this study, we explore the detectability of deuterium in B-, A-, and F-type stars, which possess radiative envelopes that can delay the mixing and destruction of recently accreted material. 
   }
   {We used synthetic spectra to generate model observations including deuterium, focusing on Balmer line regions, for stars with effective temperatures between 7\,500\,K and 12\,500\,K and a surface gravity of $\log g = 4.0$. To assess the detectability of deuterium, we employed a Markov Chain Monte Carlo framework over a range of signal-to-noise ratios between 100 and 1000. We then applied this method to observed spectra of the A9 star HD 32115 and the B9.5 star 21 Peg. 
   }
   {We show how detection limits of deuterium abundance depend on signal-to-noise ratio, effective temperature, and projected rotational velocity. For example, for a 10\,000\,K star, the detection limit decreases from D/H\,$=-4.6$ dex to $-5.5$ dex, as the signal-to-noise ratio increases from 100 to 1000. For HD 32115, we find an upper limit of D/H $< -5.5$~dex, and for 21 Peg $< -4.9$~dex. We conclude that the detection of deuterium on early-type stars may be possible in some heavily accretion-contaminated cases, providing a new diagnostic tool for the study of proto- or exo-planetary material.} 
   {}


   \maketitle

\section{Introduction}

Deuterium (D) was primarily produced during Big Bang nucleosynthesis and is destroyed inside stars. In this paper, we explore the detectability of recently accreted deuterium in the photospheres of early-type stars.

The abundance of primordial deuterium is estimated at D/H\footnote{${\rm D/H} = \log_{10}\left(N_{\rm D}/N_{\rm H}\right)$, where $N_{\rm D}$ and $N_{\rm H}$ are the number densities of deuterium and hydrogen atoms.} $=-4.59\pm0.02$ dex \citep{Cyburt2016}. Unlike nearly all other nuclei, it is easily destroyed in stellar interiors by the process of astration \citep{Epstein1976}. Subsequent stellar processing and Galactic chemical evolution modify the primordial ratio in the interstellar medium. Current estimates of the interstellar abundance range from D/H $\ge -4.70\pm0.02$ dex \citep{Prodanovic2010} to D/H $\lesssim -5$ dex \citep{Francis2025}.
For volatile molecules such as H$_{2}$O, deuterated isotopolog abundances exceeding the primordial value of D/H are commonly observed in cold astrophysical environments due to energetics favouring specific deuterated products. For example, D/H\,$=-3.77\pm0.11$ dex \citep{Cordiner2025} is measured in Solar System comets and D/H\,$=-3.82$ dex \citep{Pinti2020} in terrestrial ocean water, while protoplanetary disk gas and ices can show similarly elevated or higher levels due to low-temperature fractionation.

Despite its fragility, deuterium may be temporarily enhanced in stellar atmospheres through the accretion of deuterium-rich material, such as the engulfment of deuterium-enriched gas giants, icy bodies, or through accretion from protoplanetary or debris disks. Such episodes have the potential to increase the atmospheric deuterium abundance above primordial levels and greatly above the bulk stellar abundance.  Localised deuterium enrichment may also result from \emph{in situ} deuterium production through neutron capture by hydrogen, induced by gamma-ray irradiation from a nearby source \citep{2022OAP....35...13A}. 

However, the lifetime of any such enhancement is limited by stellar structure and mixing. If rapid convective mixing takes place, nuclear processing in stellar interiors will cause any deuterium delivered to the surface to be short-lived.
Deuterium is readily destroyed in stellar plasma once temperatures exceed $\sim 10^{6}\,\mathrm{K}$, where the $^{2}\mathrm{H}(p,\gamma)^{3}\mathrm{He}$ reaction proceeds efficiently \citep[e.g.][]{Maeder2009}. Owing to the low binding energy of deuteron (2.2~MeV), deuterium burns at significantly lower temperatures than hydrogen, so even modest convective mixing into warm interior layers leads to rapid depletion.
For deuterium to remain detectable over astrophysically meaningful timescales, it must persist in a radiative, non-convective region of the stellar envelope.

Stars with mass above $1.4\,$M$_{\odot}$, which we refer to here as early-type or BAF-type stars, provide a favourable environment for the temporary retention of deuterium. These stars have radiative envelopes that are relatively stable against convective mixing \citep[e.g.][]{Smalley2004}, allowing accreted material to remain in the photosphere for longer timescales before diffusing inward and being processed by nuclear burning. This makes BAF-type stars promising targets in the search for deuterium signatures resulting from recent accretion events.

\begin{figure*}
\centering
\includegraphics[width=0.48\textwidth]{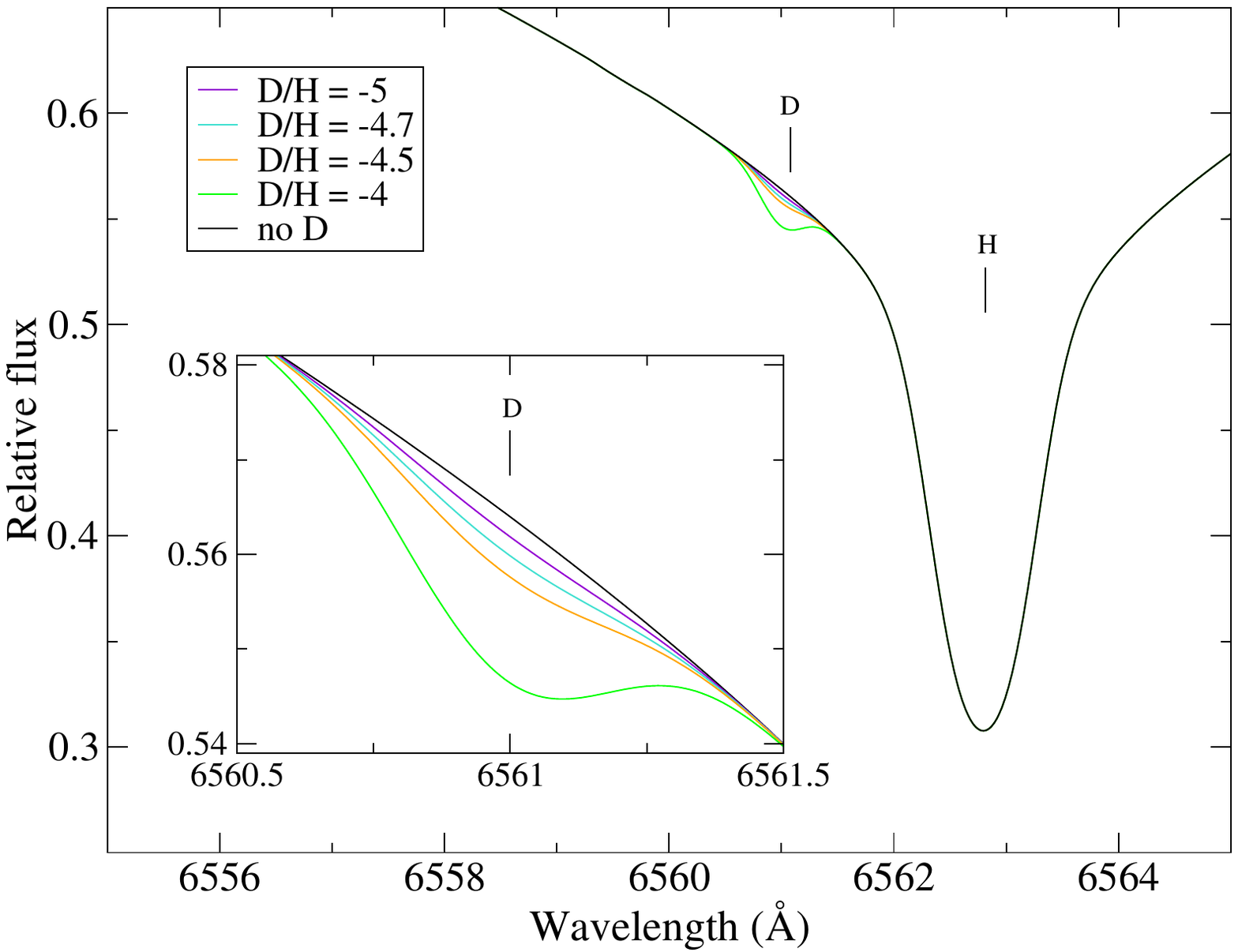}
\hfill
\includegraphics[width=0.48\textwidth]{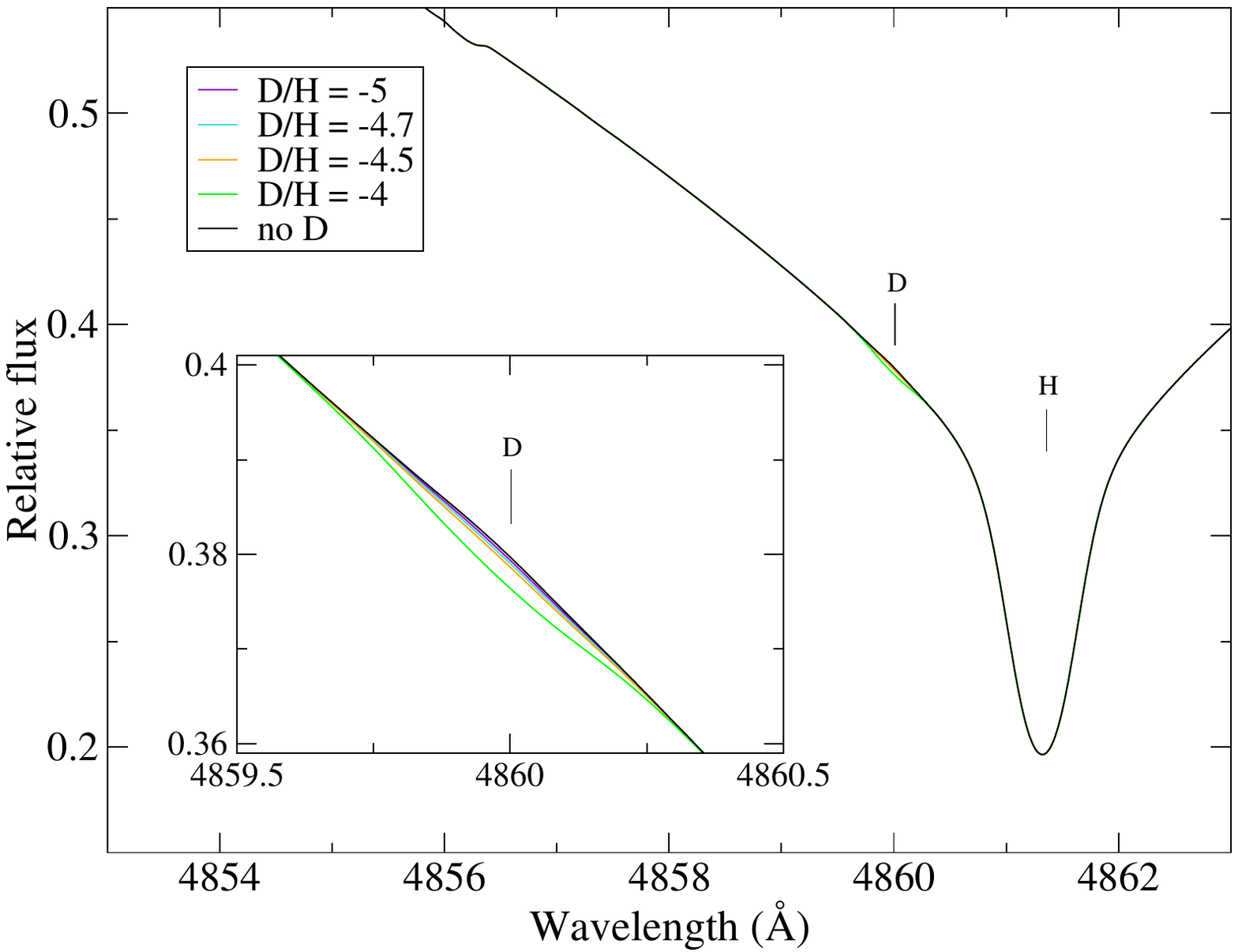}
\caption{Synthetic spectra in the H$\alpha$ region (left panel) and H$\beta$ region (right panel) for deuterium lines at different abundance levels, for a typical A-type star with $T_{\mathrm{eff}} = 10\,000$ K, $\log g = 4.0$.}
\label{fig:halpha-hbeta}
\end{figure*}

The detection of deuterium in stellar spectra remains an observational challenge. Deuterium absorption lines are typically weak and located near hydrogen lines, where they are blended with the stronger hydrogen component. Additional complications arise from line broadening caused by stellar rotation, atmospheric turbulence, and instrumental effects, all of which increase line widths. Under such conditions, the weak deuterium signature becomes indistinguishable from noise or blended structures.
Previous spectroscopic studies have consistently reported only upper limits on the deuterium abundance in stellar atmospheres. A recent analysis of the Przybylski star revealed no detectable deuterium features, establishing an upper limit of D/H $< -4$ dex \citep{Andrievsky2023}. Early searches targeting magnetic stars also resulted in non-detections, with reported D/H upper limits spanning from $-3.15$ dex down to $-4.4$ dex; for example, for $\gamma$~Equ it was constrained to D/H~$< -4.37$ dex \citep{PeimbertI_1965}. Another search for deuterium in normal stars by \citet{PeimbertII_1965} established upper limits ranging from $-3.2$ dex for the B-type star $\gamma$~Pegasi to $-4.52$ dex for the F-type star $\alpha$~Persei. High-resolution H$\alpha$ spectroscopy of Canopus provided even more stringent limits, with D/H~$< -5.04$ dex \citep{Peimbert1981} and D/H~$< -6.3$ dex \citep{Ferlet1983}, which are substantially below interstellar values. Together, these findings demonstrate the difficulty of detecting stellar deuterium and suggest that depletion, mixing, or observational limitations suppress measurable signatures.

Theoretical modelling and synthetic spectra are essential tools for estimating the detectability of deuterium features under different stellar and observational conditions. In this study, our objective is to evaluate the detectability of deuterium in the optical spectra of BAF-type stars using synthetic spectra. We focus on Balmer line regions because they offer the most favourable combination of physical sensitivity and accessibility of the spectroscopic data. Alternative hydrogen series introduce substantial complications. Paschen and Brackett lines are generally weaker, since they arise from higher energy levels with lower population densities. Some Paschen and Brackett lines are accessible from ground based spectrographs, with telluric corrections, but others (notably Pa$\alpha$) are lost in strong telluric bands. 
The wavelength separation between D and H lines is larger in the Paschen series than in the Balmer series, although most line broadening processes are also larger in wavelength for the Paschen series, partially offsetting this advantage. Lyman lines are stronger than Balmer lines, but they require space based UV observations, and are strongly affected by interstellar absorption. In some cases, circumstellar emission can also dominate these transitions, making separation of photospheric and non-photospheric contributions significantly more uncertain.  

\section{Methods}

To investigate the detectability of deuterium in BAF-type stars, synthetic spectra were generated for stellar models with effective temperatures ($T_{\mathrm{eff}}$) of 7\,500\,K, 10\,000\,K, and 12\,500\,K and a surface gravity of $\log g = 4.0$. These spectra were produced  under assumptions of local thermodynamic equilibrium (LTE) and plane-parallel geometry using the {\sc zeeman} code, a polarised radiative transfer and spectral synthesis tool \citep{Landstreet1988, Wade2001, Folsom2012}. {\sc zeeman} was extended to calculate D lines, using physics similar to H lines but accounting for differences in atomic mass. As input, a grid of {\sc atlas9} model atmospheres \citep{Kurucz1993,CastelliKurucz2004} and spectral line data extracted from the Vienna Atomic Line Database\footnote{\texttt{\href{https://vald.astro.uu.se/}{https://vald.astro.uu.se/}}} \citep[VALD;][]{Piskunov1995,Ryabchikova1997,Kupka1999,Kupka2000,Ryabchikova2015,Pakhomov2019} were used. Deuterium spectral lines, extracted from the NIST Atomic Spectra Database\footnote{\texttt{\href{https://www.nist.gov/pml/atomic-spectra-database}{https://www.nist.gov/pml/atomic-spectra-database}}} \citep{NIST_ASD}, were manually incorporated into the VALD line list. 

The partition functions for hydrogen and deuterium were evaluated explicitly for each layer of a model atmosphere.  For this, energy levels of H and D were calculated directly \citep[e.g.][Eq.\ 6.66]{Griffiths1994-quantum-intro}, accounting for the differences in nuclear mass and thus reduced mass of the system.  This produced energy levels consistent with values from NIST.  In the partition function calculation \citep[e.g.][Eq.\ 1.18]{Gray2005-Photospheres}, summation over the first 20 energy levels was used, as the inclusion of additional higher levels produces a negligible difference.  The resulting partition function values were generally consistent with the approximations from \citet{Irwin1981-partition-functions} in the 1000 -- 20\,000~K range tested.  In this range, the deuterium and hydrogen partition functions were nearly identical, so that they could be reasonably approximated with the same function, but we retained the more exact, separate values here.  

Broadening of deuterium spectral lines was modelled similarly to hydrogen lines, by incorporating thermal, Stark, and van der Waals or self-broadening effects, as well as the usual microturbulence, radial-tangential macroturbulence (kept at zero in this study), rotation, and instrumental broadening. Thermal broadening was adjusted for the different mass of the D atom \citep[e.g.][Eq.\ 11.39]{Gray2005-Photospheres}, which produces a noticeably narrower line core.  Stark broadening for D was treated identically to H, using the Vidal-Cooper-Smith model of \citet{Vidal1973}, as implemented in the tables of \citet{Lemke1997}.  Resonance self-broadening was also calculated the same way as for H using the theory of \citet{AliGriem1965, AliGriem1966}. The implementation of these Stark and self-broadening prescriptions for H lines in {\sc zeeman} code is described in \citet{Folsom2022}. This approach for Stark and self-broadening of D lines is only approximate, but should be adequate when the D abundance is low and the wings of the lines are weak. Oscillator strengths and level energies for D were taken from the NIST Atomic Spectra Database. 

\begin{figure}
  \centering
  \includegraphics[width=\hsize]{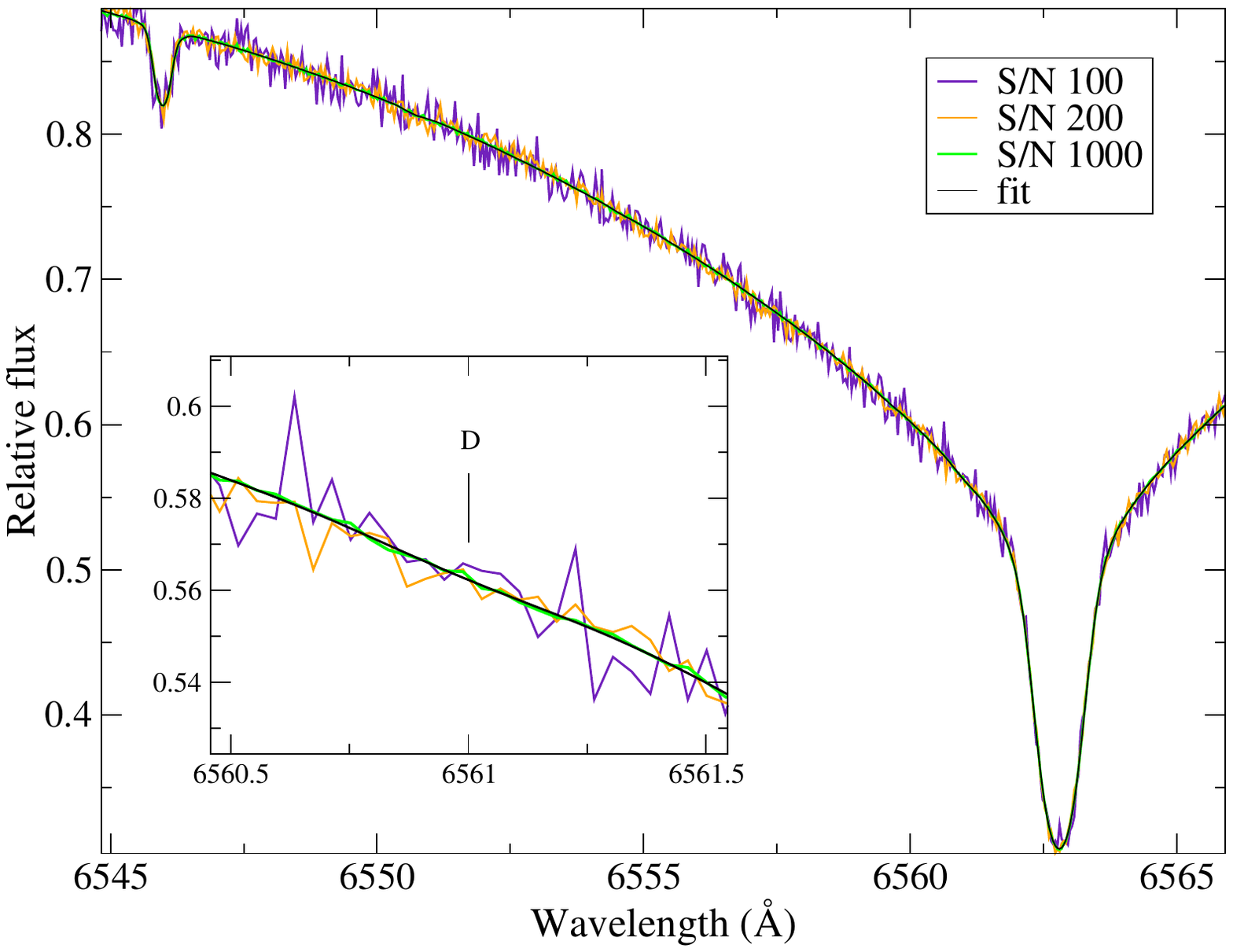}
  \caption{Synthetic spectra for \teff\ = 10\,000\,K and \logg\ = 4.0 with Gaussian noise added to simulate spectra with S/N = 100, 200, and 1000. The deuterium abundance is D/H\,=\,--5 dex.}
  \label{fig:synthD5fit}
\end{figure}

\begin{figure}[htbp]
  \centering\includegraphics[width=0.98\columnwidth]{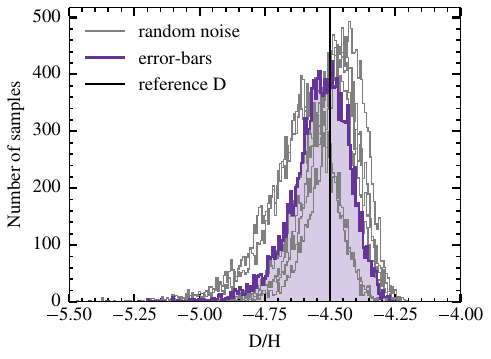}
  \caption{Distribution of the recovered D abundance for synthetic observations at \teff\ = 10\,000\,K with an input value of $-4.5$~dex (vertical line). Grey lines show results for synthetic observations with random Gaussian noise realisations, while the purple line shows the result for the deterministic error bars corresponding to the same S/N = 200.   }
  \label{fig:histograms}
\end{figure}

Figure \ref{fig:halpha-hbeta} shows synthetic spectra of the H$\alpha$ and H$\beta$ regions for a typical A-type star ($T_{\mathrm{eff}} = 10\,000$ K, $\log g = 4.0$) at different deuterium abundances. The deuterium feature is much more pronounced in H$\alpha$ than in H$\beta$, due to the different oscillator strengths, illustrating that H$\alpha$ provides a more sensitive diagnostic for measuring deuterium in these stellar atmospheres.

We constrained deuterium abundances in synthetic observed spectra using a Markov Chain Monte Carlo (MCMC) approach implemented with the {\sc ZMCwrap} tool\footnote{\texttt{\href{https://github.com/folsomcp/ZMCwrap}{https://github.com/folsomcp/ZMCwrap}}} (Folsom et al., in prep.). {\sc ZMCwrap} is built as a Python wrapper around the {\sc zeeman} spectrum synthesis code, and it uses the {\sc emcee} package \citep{Foreman-Mackey2013} to run the MCMC sampler. We adopt separable uniform priors, subject to motivated constraints based on physical and numerical limits. Rotational and turbulent velocities are required to be non-negative. Effective temperature \teff\ is restricted to $3500 - 30\,000$~K and surface gravity $\log{g}$ to $2.5 - 5.0$. The deuterium abundance is constrained to $-13.0 \le \mathrm{D}/\mathrm{H} \le -0.5$, for numerical reliability in the spectrum synthesis code. We include a multiplicative continuum correction to account for residual normalisation errors across the fitted wavelength interval. The model flux is scaled by a low-order Chebyshev polynomial expansion, with coefficients treated as free parameters. An additional free parameter $\log f_\sigma$ was included, which represents the logarithm of a global scaling factor applied to the observational uncertainties, \citep[following e.g.][]{Hogg2010-fitting-data, Anfinogentov2021-SoBAT-MCMC}. This parameter allows for accounting for possible misestimation of the noise level in the data and ensures statistically consistent uncertainty estimates. The $\log f_\sigma$ parameter is optimised as part of the MCMC analysis, and marginalised over for the resulting D/H posterior distribution. In practice, this has the result of increasing the uncertainty on D/H, in a fashion that is consistent with hypothesis that the scatter in the data about the optimal model is driven by Gaussian noise.

This statistical method allowed us to retrieve the posterior probability distributions of the model parameters, including the deuterium abundance, in synthetic observations and real observed spectra.
To visualise the marginalised posterior distributions and assess parameter correlations, we used the {\sc corner} package \citep{Foreman-Mackey2016}. This toolkit produces corner plots (also known as triangle plots), which provide a representation of the one- and two-dimensional projections of the posterior distributions. These plots were used to assess the statistical significance of deuterium detection.

\section{Analysis}

Our analysis consisted of two parts. The first is a set of synthetic retrieval tests, using the code both as a forward model and as a retrieval model, to verify our methodology and determine theoretical limits on sensitivity to deuterium. The second is an MCMC analysis of real observations of two well-studied stars.

\subsection{Synthetic retrievals}

We calculated synthetic spectra with known model parameters, and used them to produce synthetic observations for the MCMC analysis to test constraints on D abundances. 
The synthetic spectra were generated assuming fixed stellar and broadening parameters, representative of a typical slowly rotating intermediate-mass star. 
The adopted values were radial velocity $V_r = 0\ \mathrm{km\ s^{-1}}$, projected rotational velocity $v \sin i = 10\ \mathrm{km\ s^{-1}}$, microturbulent velocity $v_{\mathrm{mic}} = 1\ \mathrm{km\ s^{-1}}$, macroturbulent velocity $v_{\mathrm{mac}} = 0\ \mathrm{km\ s^{-1}}$, surface gravity $\log g = 4.0$ and solar metallicity. Three effective temperatures \teff\ were considered 7\,500\,K, 10\,000\,K, and 12\,500\,K. A range of D abundances were tested: no deuterium, and from D/H~$=-5.5$~dex to $-4$~dex in steps of 0.5 dex.

The study examined the detectability of deuterium in the H$\alpha$, H$\beta$ and H$\gamma$ regions. 
To emulate realistic observations, we convolved the synthetic spectra with a Gaussian instrumental profile at a resolution of 65\,000, and then resampled the spectra onto a grid of 1.8 {km\,s$^{-1}$} pixels. 
We tested two approaches to simulate the uncertainties in the observed data points. In the first approach, Gaussian noise was added, scaled with the continuum flux to match a user-defined peak S/N. Figure\,\ref{fig:synthD5fit} illustrates some of these model observations with different S/N levels. However, we found that the recovered distributions of the elemental abundances depended on the specific noise realisation. In particular, the peak of the resulting histograms did not coincide with the input value used in the synthetic spectrum. This effect is consistent with expectations from the input noise, as the true input value is recovered within the 16th–84th percentile interval about 68\% of the time. However, obtaining unbiased sensitivity limits would require averaging results across many independent MCMC runs. Figure\,\ref{fig:histograms} demonstrates that different noise realisations produce histograms with peaks shifted relative to the input value. 
In contrast, when only wavelength-dependent error bars are provided, the recovered distribution peaks at the correct value of $-4.5$~dex. To avoid biases introduced by stochastic noise  and to ensure that the recovered distributions accurately reflect the input parameters, we therefore adopted the deterministic approach using error bars only. 

\begin{figure}
  \centering
  \includegraphics[width=\hsize]{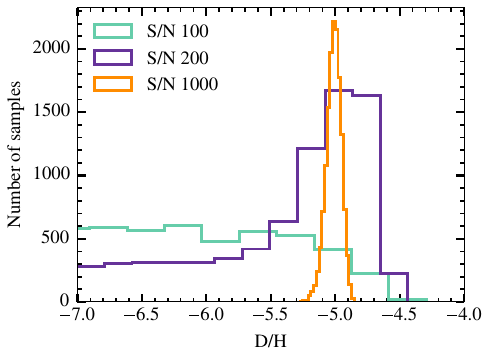}
  \caption{MCMC-derived posterior histograms illustrating deuterium recovery for a $T_{\mathrm{eff}} = 10\,000\ \mathrm{K}$ star under varying noise conditions: detection (S/N\,$=1000$), marginal detection (S/N\,$=200$), and non-detection (S/N\,$=100$) for an input value D/H\,=\,--5 dex.}
  \label{fig:D5}
\end{figure}

\begin{figure}
  \centering
  \includegraphics[width=\hsize]{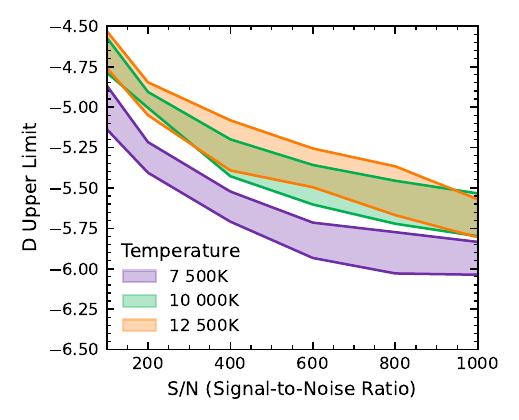}
  \caption{Detection limit of deuterium as a function of S/N for three different \teff\ values. The shaded region is bounded by two contours: the upper edge corresponds to the 99.9\% confidence level and the lower edge to the 99\%.}
  \label{fig:upperlimit}
\end{figure}

\begin{figure}[htbp]
  \centering\includegraphics[width=0.98\columnwidth]{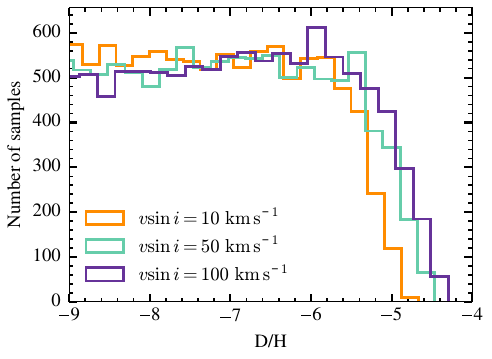}
  \caption{Posterior probability distributions of D/H for different \vsini, for a model with \teff\ =\,10\,000\,K.}
  \label{fig:vrot}
\end{figure}

Figure\,\ref{fig:D5} shows histograms of the posterior distribution of the deuterium abundance obtained from MCMC runs on synthetic observations with different S/N values. All MCMC samplings were performed with 50 walkers ($n_{\mathrm{walkers}})$ over 600 steps ($n_{\mathrm{chain}}$), discarding the first 300 steps as burn-in ($n_{\mathrm{burn}}$). 
The input synthetic spectra assume D/H$=-5$~dex, with fixed stellar parameters ($T_{\mathrm{eff}} = 10\,000\ \mathrm{K}$, $\log g = 4.0$, $V_r = 0\ \mathrm{km\ s^{-1}}$, $v \sin i = 1\ \mathrm{km\ s^{-1}}$, $v_{\mathrm{mic}} = 10\ \mathrm{km\ s^{-1}}$, and solar metallicity).
A high sensitivity, S/N\,$=1000$, yields a narrow posterior centred near the true input value (D/H$=-5$~dex), while an intermediate sensitivity of S/N\,$=200$ results in a broader distribution with an extended tail to very small values, indicating that this is a marginal detection, where the presence of D is likely but not certain. At low S/N = 100, the posterior lacks a pronounced peak and remains largely flat down to very low abundance values, denoting insufficient information for a reliable detection. 

Figure\,\ref{fig:upperlimit} illustrates the dependence of the inferred detection limits on  D/H as a function of S/N, derived from the MCMC analysis of stellar models with $T_{\mathrm{eff}}$ = 7\,500\,K, 10\,000\,K, and 12\,500\,K. For each temperature, the shaded regions represent the range between the 99.0\% and 99.9\% confidence limits, thereby quantifying the statistical uncertainty associated with the inferred deuterium abundance constraints. The corresponding values are listed in Table \ref{table:1}.

A clear and systematic trend is observed across all models: increasing S/N leads to progressively tighter constraints on D/H. Over the explored range, the detection limits decrease approximately monotonically with S/N, demonstrating the dominant role of data quality in deuterium detectability. At S/N\,$\approx$\,100, the constraints are relatively weak, whereas at S/N\,$\geq$\,800 the limit converges towards asymptotic values.

The 7\,500\,K model consistently yields the lowest detection limits at all S/N values, indicating that cooler early-type stars provide the most favourable conditions for deuterium detection. In contrast, the 12\,500\,K model exhibits systematically higher detection limits, reflecting reduced  sensitivity. 

To examine how the detection limit on deuterium abundance depends on rotational broadening, we performed MCMC simulations using synthetic spectra with no deuterium included. The spectra were generated for a stellar atmosphere with \teff~$=10\,000$~K and S/N $=200$. The analysis was performed in the wavelength range 6460–6668 \AA\, around the H$\alpha$ line.
Three projected rotational velocities were considered: $v \sin i = 10, 50$ and $100\,\mathrm{km\,s^{-1}}$

 The posterior distributions of D/H for different projected rotational velocities are compared in Figure\,\ref{fig:vrot}. The smaller detection limits correspond to the smaller $v \sin i$ of $10\,\mathrm{km\,s^{-1}}$, indicating a good sensitivity to D abundance. The loss of sensitivity is most pronounced between $v \sin i = 10$ and $50\,\mathrm{km\,s^{-1}}$, while the degradation between $50$ and $100\,\mathrm{km\,s^{-1}}$ is comparatively smaller. The velocity separation between the D$\alpha$ and H$\alpha$ lines is relatively small, only $\approx 81\,\mathrm{km\ s^{-1}}$, implying that once $v \sin i$ reaches this value, the line cores become heavily blended. For this reason, it is preferable to consider objects with $v \sin i\lesssim 40\,\mathrm{km\ s^{-1}}$ to mitigate blending and reduce sensitivity to systematic errors in the H$\alpha$ line core model. The 99th percentile detection limits are ${\rm D/H} = -5.13$, $-4.86$, and $-4.70$ dex for $v \sin i = 10$, $50$, and $100\,\mathrm{km\,s^{-1}}$, respectively.

Following the analysis of synthetic observations, the method was applied to real stellar spectra.

\begin{table}
\caption{Estimated limits of deuterium detection for stellar temperatures \teff\ 7\,500\,K, 10\,000\,K, and 12\,500\,K
 across a range of S/N from 100 to 1000 , given at 99\% and 99.9\% confidence levels.}
  \label{table:1}
\centering
\begin{tabular}{@{}lcccccc@{}}
\hline\hline
\teff\ &\multicolumn{2}{c}{7\,500\,K}&\multicolumn{2}{c}{10\,000\,K}&\multicolumn{2}{c}{12\,500\,K}\\
\hline
Confidence    &    99       &99.9    &    99       &99.9&    99       &99.9\\
\hline
S/N 100 &   $-5.14$     &$-4.87$    &   $-4.79$     &$-4.57$ &   $-4.76$     &$-4.53$\\
S/N 200 &   $-5.41$     &$-5.22$    &   $-5.00$     &$-4.91$ &   $-5.05$     &$-4.85$\\
S/N 400 &   $-5.71$     &$-5.52$    &   $-5.43$     &$-5.20$ &   $-5.39$     &$-5.08$\\
S/N 600 &   $-5.93$     &$-5.71$    &   $-5.60$     &$-5.36$ &   $-5.50$     &$-5.26$\\
S/N 800 &   $-6.03$     &$-5.77$    &   $-5.72$     &$-5.46$ &   $-5.67$     &$-5.36$\\
S/N 1000 &  $-6.04$     &$-5.83$    &   $-5.80$     &$-5.53$ &   $-5.81$     &$-5.57$\\
\hline
\end{tabular}
\end{table}

\subsection{Observations of 21\,Peg and HD\,32115}

To test the method on real stars, we selected the A9 star HD\,32115 and the B9.5 star 21\,Peg, and again focused on Balmer lines.
All spectroscopic observations used in this work are archival and were accessed through PolarBase \citep{PolarBase}. 
We analysed spectra of HD\,32115 (2005-12-17) and of 21\,Peg (2013-08-17, observed in polarimetric mode) obtained with ESPaDOnS.
ESPaDOnS is a cross-dispersed echelle spectrograph mounted on the 3.6\,m Canada--France--Hawaii Telescope (CFHT) on Mauna Kea, Hawaii, delivering a spectral resolution of about $R \sim 65{,}000$ over a wavelength range of 3\,700--10\,500\,\AA\ \citep{Donati2003}.
The S/N of the spectra in the H$\alpha$ wavelength region reaches $\approx150$ for HD\,32115 and $\approx840$ for 21\,Peg. The total exposure times are 60 s and 1160 s, respectively.

We adopted the atmospheric parameters for HD\,32115 from \citet{Fossati2011}: $T_{\mathrm{eff}} = 7\,250 \pm 100$\,K, $\log g = 4.2 \pm 0.1$, $v_{\mathrm{mic}} = 2.5 \pm 0.2$\,km s$^{-1}$, $v \sin i = 8.3\,\mathrm{km\ s^{-1}}$. Similarly, for 21\,Peg, we used the parameters from \citet{Fossati2009}: $T_{\mathrm{eff}} = 10\,400 \pm 200$\,K, $\log g = 3.5 \pm 0.1$, $v_{\mathrm{mic}} = 0.5 \pm 0.4$\,km\,s$^{-1}$, $v \sin i = 3.76\pm 0.35$\,km s$^{-1}$, and $V_r = 0.5 \pm 0.5$\,km\,s$^{-1}$.
The authors derived the fundamental parameters using multicolour photometry, pressure-sensitive magnesium lines, metallic lines, and Balmer line profiles. $T_{\mathrm{eff}}$ and $\log{g}$ were refined from photometric estimates via Balmer line fitting and Mg\,{\sc i} wings. The microturbulent velocity $v_{\mathrm{mic}}$ was derived by minimising the correlation between the abundances of metal-lines and the equivalent widths. 

To constrain the deuterium abundance, we ran the MCMC retrieval with seven free parameters: the stellar effective temperature and surface gravity, three continuum polynomial coefficients (C1, C2, and C3), $\log f_\sigma$, and the deuterium abundance. Here we used the same priors as described in Sect. 2, and uniform priors in C1, C2, C3, and $\log f_{\sigma}$. The analysis focused on a narrow region around the D line. The H line core was excluded from the fit to avoid any potential bias from non-LTE effects in the line core. As a consequence, the spectral regions did not contain sufficient information to reliably constrain $T_{\mathrm{eff}}$ and $\log{g}$.  However, they were included in the fit, and treated as nuisance parameters along with the continuum polynomial, to provide the best fit possible around the D line and avoid biasing the D abundance determination. All MCMC samplings were performed with $n_{\mathrm{walkers}}=100$, $n_{\mathrm{chain}}=1400$, $n_{\mathrm{burn}}=700$.

\begin{figure}
    \centering 
    \includegraphics[width=\linewidth]{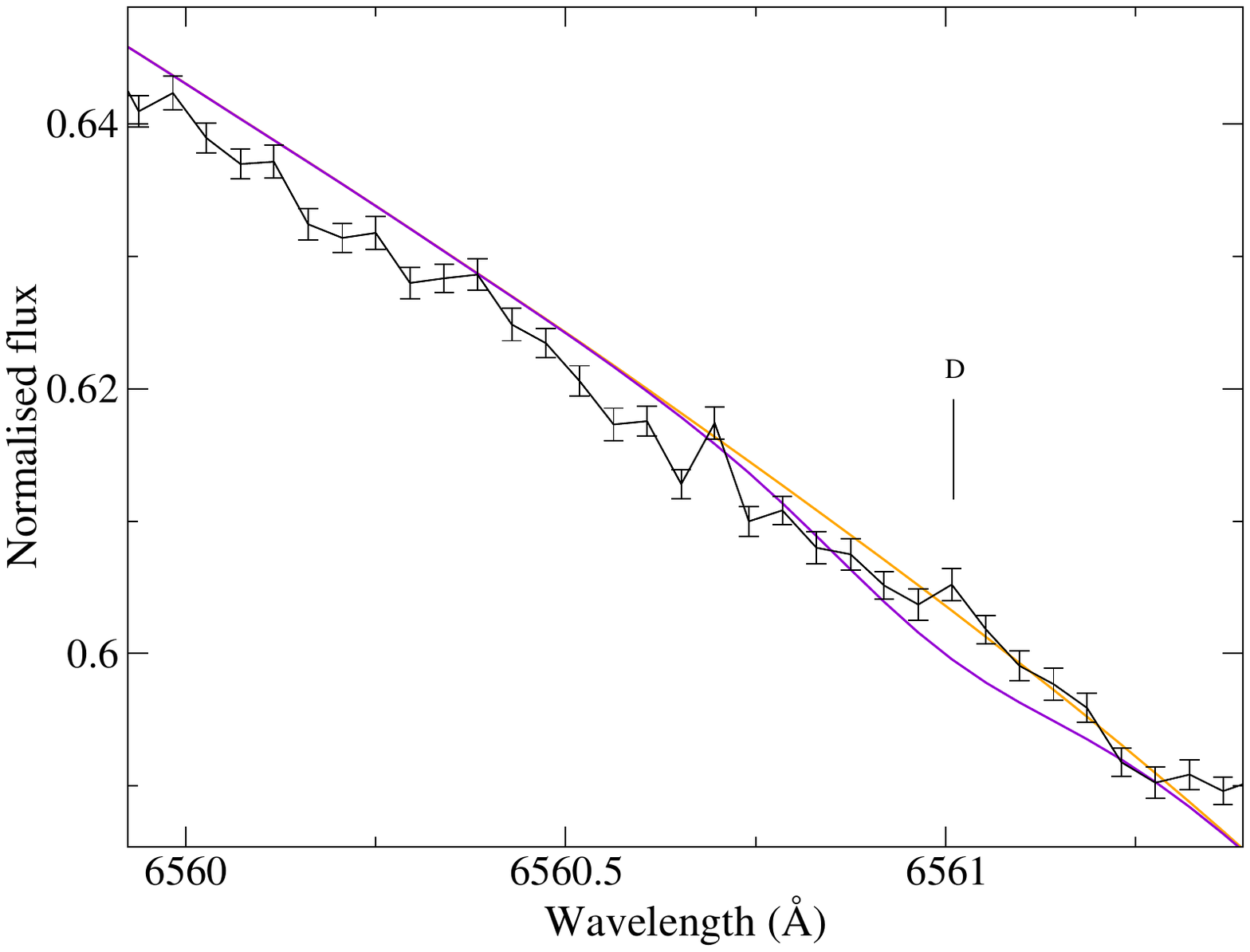}
    \includegraphics[width=\linewidth]{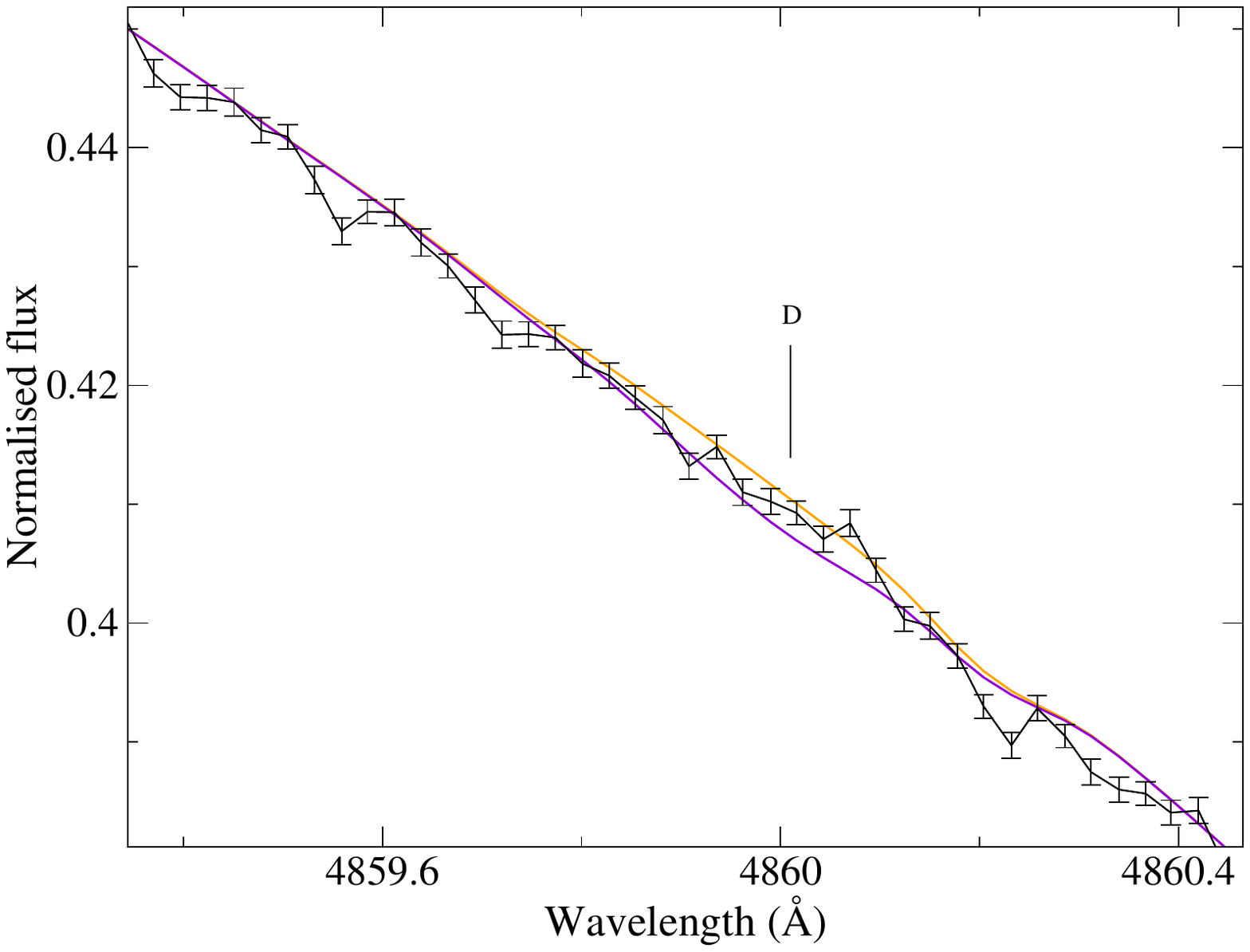}
    \caption{Comparison of the observed spectrum of 21\,Peg (black) in the H$\alpha$ (top) and H$\beta$ (bottom) regions with the best-fit (orange) and the upper-limit (purple) models of the D line. In the H$\alpha$ region, the best-fit and the upper-limit models correspond to D/H\,=\,$-8.92$ and $-4.70$\,dex, respectively. In the H$\beta$ region, the corresponding values are D/H\,=\,$-8.06$ and $-4.53$\,dex. }
    \label{fig:21Peg_bestfit}
\end{figure}

21\,Peg: The model fitting for this star was restricted to two spectral intervals 6553.0–6561.7 \AA\ and 4857.0–4860.66 \AA. Figure\,\ref{fig:21Peg_bestfit} displays resulting fits for corresponding wavelength regions, with the best fit parameters taken from the median values of the MCMC chains. We show the posterior distributions of the stellar parameters inferred from fitting the H$\alpha$ (Figure\,\ref{fig:21PegHaCorner}) and H$\beta$ (Figure\,\ref{fig:21PegHbCorner}) line profiles of 21\,Peg. The two-dimensional marginalised posterior reveal strong correlations between $T_{\mathrm{eff}}$, $\log{g}$ and the continuum polynomial coefficients, in part due to the limited wavelength range used. However, the D abundance does not strongly correlate with any of these parameters. We derive upper limits of D/H $\leq-4.89$\,dex at 99th percentile and $-4.64$ dex at $99.9$th percentile for D$\alpha$. For D$\beta$, the corresponding upper limits are D/H $\leq-4.71$\,dex (99th percentile) and $-4.50$\,dex ($99.9$th percentile).

\begin{figure}
  \centering
  \includegraphics[width=0.98\columnwidth]{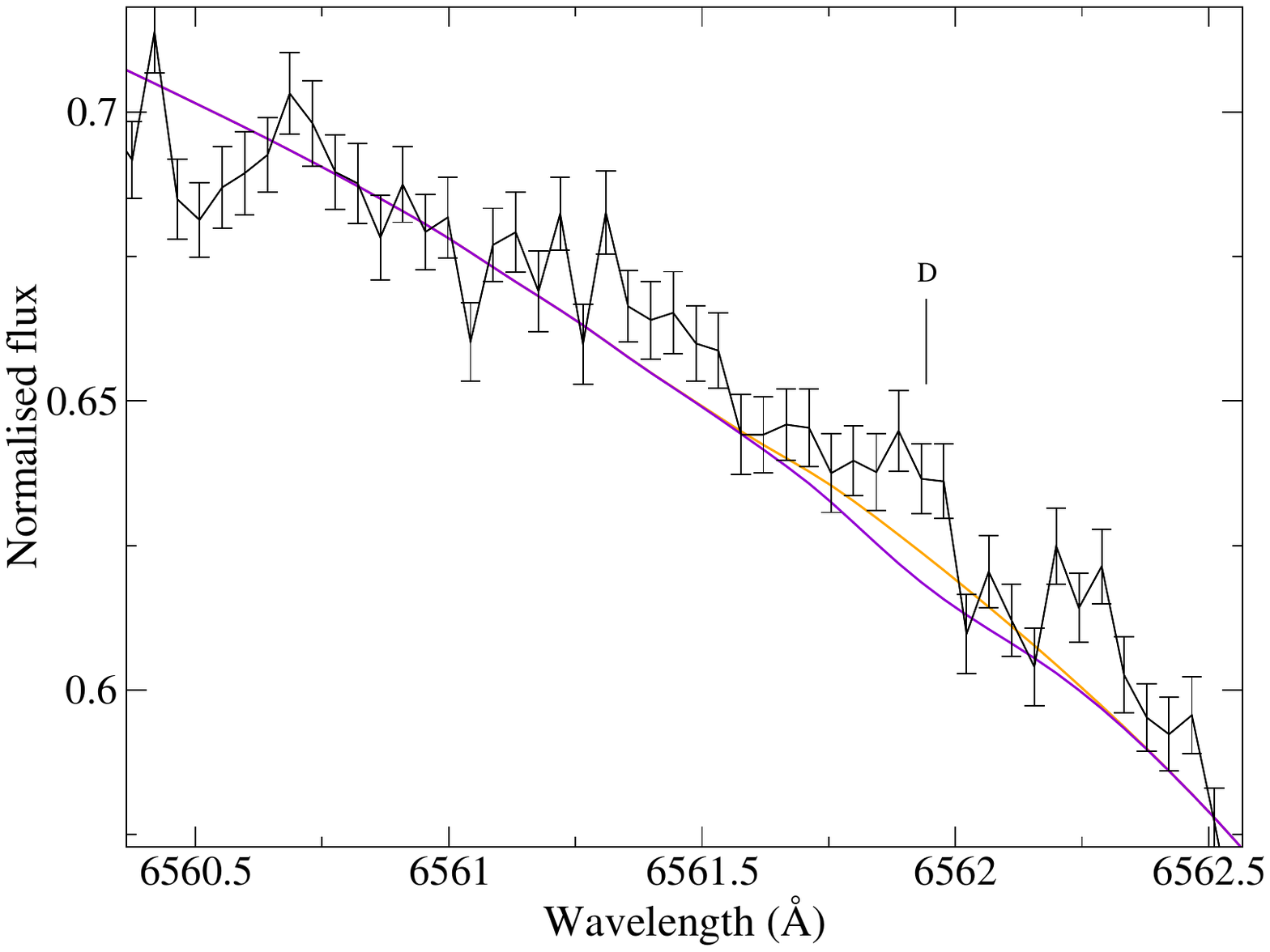}
  \caption{The best-fit (orange) and the upper-limit (purple) models of the D line compared with the observed spectrum of HD\,32115 (black) in H$\alpha$ wavelength region. D/H is $-9.41$ dex for the best-fit model and $-5.29$ dex for the upper-limit.}
  \label{fig:HD32115Hafit}
\end{figure}

HD\,32115: The analysis for this star was limited to the wavelength window 6556.6–6562.7\,\AA. The radial velocity $V_r$ was fixed to $43\,\mathrm{km\,s^{-1}}$, a value derived from prior $\chi^2$ fitting of metal lines in the 5\,000\,--\,5\,100 \AA\ region, using the {\sc zeeman} code. The posterior distributions are shown in Figure\,\ref{fig:HD32115Ha} and the best-fit spectrum is shown in Figure\,\ref{fig:HD32115Hafit}. The posterior distribution again shows strong correlations between $T_{\mathrm{eff}}$, $\log{g}$ and the continuum polynomial coefficients, this time with $\log{g}$ largely unconstrained from portion of the H$\alpha$ wing.  However, the D abundance is again uncorrelated with the nuisance parameters, implying that this value is largely uneffected by the relatively poor constraint on those other values. Upper limits of D/H $\leq-5.53$\,dex at 99th percentile and D/H $\leq-5.12$ dex at $99.9$th percentile are obtained.

\section{Discussion}

We have demonstrated that for a well-normalised spectrum of a slowly rotating early-type star with S/N\,$\gtrsim$\,200, the deuterium abundance in the photosphere can be constrained to D/H\,$\lesssim-5$\,dex.
Below, we discuss the results of testing our method on two stars, and explore the usefulness of such a constraint for different scenarios of accreting material.

\subsection{Results for 21\,Peg and HD\,32115}
Upper limits on the photospheric deuterium content were obtained for both stars. The non-detections are consistent with the relatively mature ages, $\approx 268\,$Myr for 21\,Peg and $\approx 780\,$Myr for HD\,32115 \citep{GAIADR3}. These ages have given enough time for diffusive and rotational mixing to cycle any surface-level deuterium through the D-burning ($T\sim10^{5}$\,K) layers. A further conclusion is that there is no significant current or recent accretion of material with an interstellar-like (protoplanetary or jovian) deuterium abundance. This is consistent with the approximately solar photospheric abundances found for most elements in these stars by \citet{Fossati2009} and \citet{Fossati2011}.

\subsection{Deuterium enhancement from accretion}

In intermediate-mass stars, any deuterium inherited from their birth environment during initial assembly of the star is burned during the fully convective part of the pre-main sequence evolution, once temperatures reach $\gtrsim10^{6}\,$K \citep{Stahler1988}. Once the radiative envelope is established, the deuterium-depleted core becomes decoupled from the envelope and thin convective surface, which can be ``polluted'' with subsequent layers of accretion from a protoplanetary disk, or potentially gas-rich debris disks or even evaporating planets \citep{JermynKama2018}. Given enough time, rotational and diffusive mixing will transport any newly-accreted D into the $\gtrsim10^{6}\,$K zone of the star, where it will be burned. We do not expect to find deuterium even at the baseline cosmic abundance in the photospheres of main-sequence intermediate-mass stars, unless it has been accreted there from some circumstellar reservoir within approximately one diffusive or rotational mixing timescale, whichever is fastest, from the present. Therefore, even a detection of a nominal interstellar D/H value would be informative, indicating relatively recent accretion. This is even more strongly the case for any non-cosmic D/H abundance. We discuss some of the possible sources of accreted D below.

Protoplanetary disks around Herbig\,Ae/Be stars typically feed the star at rates $\dot{M}\sim10^{-8}$ to $10^{-6}$\,M$_{\odot}$\,yr$^{-1}$ \citep[e.g.,][]{Mendigutiaetal2012, Fairlambetal2015}. 
The fraction of accreted material mixed into the stellar photosphere, $f_{\rm ph}$, increases with accretion rate. For accretion rates near the upper end of this range, the photospheric material becomes almost entirely dominated by accreted matter, $f_{\rm ph}\approx1$, even for very fast rotation \citep{JermynKama2018}.  
For $\dot{M}$ at the lower end, near $10^{-8}\,$M$_{\odot}$\,yr$^{-1}$, the mixing fraction of accreted material remains at $f_{\rm ph}\sim1$ up to rotational velocities of $v_{\rm rot}\approx100\,$km\,s$^{-1}$ for $6\,$M$_{\odot}$ stars, and up to $\approx10\,$km\,s$^{-1}$ for $1.5\,$M$_{\odot}$ stars. A nominal interstellar deuterium abundance of D/H$= -4.7\pm0.02$ dex should therefore be detectable on their surface at S/N\,$\gtrsim200$ (see Figure\,\ref{fig:upperlimit}), assuming an otherwise good fit can be obtained to the spectrum around the D line.

Some evidence for accretion contamination influencing photospheric abundances has been observed in some protoplanetary disk stars.  In particular, photospheric depletions of refractory elements \citep{Folsom2012} are related to circumstellar disk structure, and such depletions appear to be a consequence of a recent accretion of refractory-depleted material \citep{Kama2015, JermynKama2018}.  Further studies support this hypothesis \citep{Kama2016, Murphy2021, GuzmanDiaz2023, Borthakur2025}, strengthening the possibility of detecting recently accreted D in the photosphere of some of these stars. 

Gas-rich debris disks are predicted to have gas accretion rates onto their host stars of as high as $\dot{M}\sim10^{-11}$\,M$_{\odot}$\,yr$^{-1}$, though this may be exceptional and the high-end numbers may usually be closer to $10^{-14}$--$10^{-13}$\,M$_{\odot}$\,yr$^{-1}$ \citep{Kraletal2017, Kraletal2019}. With such accretion rate values, the steady-state mixing fraction, $f_{\rm ph}$, of fresh material in early-type star photospheres will be low. At best, $f_{\rm ph}\sim 10^{-2}$--$10^{-1}$, and likely closer to $10^{-4}$--$10^{-3}$ \citep{JermynKama2018}. This unfavourable mixing ratio may be partly offset by the potential for the secondary gas in some of these disks to be dominated by CO, CO$_{2}$, H$_{2}$O, and other abundant ices. In such a scenario, the deuteration fraction of the gas may represent the net deuteration fraction of water ice, (HDO$+$D$_{2}$O)/H$_{2}$O, in protoplanetary disks. Depending on the chemical inheritance of protoplanetary water, this ratio may be in the range from a few times above the cosmic D/H value to perhaps as high as D/H\,$\sim0.1$--$1\,$\% \citep[e.g.,][]{Coutens2012, Tobin2023, Slavicinska2025, Leemker2025}. Due to the low overall accretion rates, the likelihood of detecting photospheric deuterium accreted from gas-rich debris disks appears to be low.

Exoplanets may transfer mass onto the stellar photosphere either through atmospheric escape and subsequent accretion, or by being engulfed by the star. Giant planet atmospheric mass loss rates due to UV irradiation are thought to be in the range $\dot{M}=10^{-16}$--$10^{-13}\,$M$_{\odot}$\,yr$^{-1}$ \citep{Owen2019, Wyttenbach2020, Krenn2021}. Even assuming perfect efficiency of mass transfer onto the star, which is unlikely, such rates are insufficient to generate a measurable composition signature on the surface of an intermediate mass star \citep{JermynKama2018}. As for engulfment, the process lasts $\sim10^{3}\,$years for even low mass planets, during which the planet disintegrates as it falls deeper into the star and may leave only a small part of its mass in the upper convective layer and photosphere \citep{Lane2026}. We therefore refrain from making specific predictions, aside from stating that the D/H of a giant planet can probably be expected to be within a factor of $1$ to $10$ of the cosmic D/H ratio, as is the case in the case for the solar system gas and ice giants \citep{Feuchtgruber1999, Lellouch2001, Pierel2017}.

\section{Conclusions}

This work provides a foundation for future observational efforts to search for deuterium as a potential tracer of recent accretion events in early-type stellar atmospheres. We have assessed the detectability of deuterium in the spectra of early-type stars for a range of realistic conditions. 

The analysis shows that the deuterium line is more prominent in the  H$\alpha$ region compared to  H$\beta$ or H$\gamma$ at the same abundance levels. This difference likely arises due to the oscillator strength of the transitions, which influences the line intensity.
Our MCMC simulations yielded the following findings: 

\begin{itemize}
        \item The sensitivity of detection depends on stellar \teff\, with cooler stars providing better constraints on deuterium abundance. For a star with \teff = 7\,500\,K, $\log g = 4.0$ and $v \sin i = 10\ \mathrm{km\ s^{-1}}$, D/H $\geq -5.0$ dex is detectable with S/N\,$\geq$\,200. For hotter stars with \teff\ $\geq$\, 10\,000\,K, S/N\,$\geq400$ is required.
        \item As the S/N increases, the detection limit of deuterium decreases, allowing for more precise constraints. Increasing the S/N from 100 to 1000 improves the detection limit by approximately one order of magnitude ($\sim$1 dex) in abundance, however, increasing the S/N above approximately 800 does not significantly improve the achievable detection limits.
        \item The sensitivity of detection also depends on $v\sin i$, with the detection limit increasing by 0.4 dex as $v\sin i$ increases from 10 to 50 km\,s$^{-1}$.
\end{itemize}

We applied the method to observational spectra of two stars in the $T_{\mathrm{eff}}$ = 7\,500 -- 10\,000\,K range, where the convective mixing is weak and our method is most sensitive. We found upper limits of D/H $\leq-4.9$\,dex for 21\,Peg and D/H $\leq-5.5$\,dex for HD\,32115 at 99\% confidence levels.

\begin{acknowledgements}
This work has received funding from the European Union's Horizon Europe research and innovation programme under grant agreement No. 101079231 (EXOHOST), and from UK Research and Innovation (UKRI) under the UK government's Horizon Europe funding guarantee (grant number 10051045). AA acknowledges support from the Estonian Research Council grant PRG 2159. 
      This work has made use of the VALD database, operated at Uppsala University, the Institute of Astronomy RAS in Moscow, and the University of Vienna.
\end{acknowledgements}

\bibliographystyle{aa}
\bibliography{references}

@ARTICLE{Cyburt2016,
       author = {{Cyburt}, Richard H. and {Fields}, Brian D. and {Olive}, Keith A. and {Yeh}, Tsung-Han},
        title = "{Big bang nucleosynthesis: Present status}",
      journal = {Reviews of Modern Physics},
         year = 2016,
        month = jan,
       volume = {88},
       number = {1},
          eid = {015004},
        pages = {015004},
          doi = {10.1103/RevModPhys.88.015004},
archivePrefix = {arXiv},
       eprint = {1505.01076},
 primaryClass = {astro-ph.CO},
       adsurl = {https://ui.adsabs.harvard.edu/abs/2016RvMP...88a5004C}
}

@ARTICLE{Stahler1988,
       author = {{Stahler}, Steven W.},
        title = "{Deuterium and the Stellar Birthline}",
      journal = {\apj},
         year = 1988,
        month = sep,
       volume = {332},
        pages = {804},
          doi = {10.1086/166694},
       adsurl = {https://ui.adsabs.harvard.edu/abs/1988ApJ...332..804S}
}

@ARTICLE{Epstein1976,
       author = {{Epstein}, R.~I. and {Lattimer}, J.~M. and {Schramm}, D.~N.},
        title = "{The origin of deuterium}",
      journal = {\nat},
         year = 1976,
        month = sep,
       volume = {263},
        pages = {198-202},
          doi = {10.1038/263198a0},
       adsurl = {https://ui.adsabs.harvard.edu/abs/1976Natur.263..198E}
}

@ARTICLE{Prodanovic2010,
       author = {{Prodanovi{\'c}}, Tijana and {Steigman}, Gary and {Fields}, Brian D.},
        title = "{The deuterium abundance in the local interstellar medium}",
      journal = {\mnras},
         year = 2010,
        month = aug,
       volume = {406},
       number = {2},
        pages = {1108-1115},
          doi = {10.1111/j.1365-2966.2010.16734.x},
archivePrefix = {arXiv},
       eprint = {0910.4961},
 primaryClass = {astro-ph.GA},
       adsurl = {https://ui.adsabs.harvard.edu/abs/2010MNRAS.406.1108P}
}

@ARTICLE{Francis2025,
       author = {{Francis}, L. and {van Dishoeck}, E.~F. and {Caratti o Garatti}, A. and {van Gelder}, M.~L. and {Gieser}, C. and {Beuther}, H. and {Ray}, T.~P. and {Tychoniec}, L. and {Nazari}, P. and {Reyes}, S. and et al.},
        title = "{JOYS: The [D/H] abundance derived from protostellar outflows across the Galactic disk measured with JWST}",
      journal = {\aap},
         year = 2025,
        month = feb,
       volume = {694},
          eid = {A174},
        pages = {A174},
          doi = {10.1051/0004-6361/202451629},
archivePrefix = {arXiv},
       eprint = {2501.02085},
 primaryClass = {astro-ph.GA},
       adsurl = {https://ui.adsabs.harvard.edu/abs/2025A&A...694A.174F}
}

@ARTICLE{Cordiner2025,
       author = {{Cordiner}, M.~A. and {Gibb}, E.~L. and {Kisiel}, Z. and {Roth}, N.~X. and {Biver}, N. and {Bockel{\'e}e-Morvan}, D. and {Boissier}, J. and {Bonev}, B.~P. and {Charnley}, S.~B. and {Coulson}, I.~M. and {Crovisier}, J. and {Drozdovskaya}, M.~N. and {Furuya}, K. and {Jin}, M. and {Kuan}, Y.-J. and {Lippi}, M. and {Lis}, D.~C. and {Milam}, S.~N. and {Opitom}, C. and {Qi}, C. and {Remijan}, A.~J.},
        title = "{A D/H ratio consistent with Earth's water in Halley-type comet 12P from ALMA HDO mapping}",
      journal = {Nature Astronomy},
         year = 2025,
        month = aug,
       volume = {9},
        pages = {1476-1485},
          doi = {10.1038/s41550-025-02614-7},
archivePrefix = {arXiv},
       eprint = {2508.05925},
 primaryClass = {astro-ph.EP},
       adsurl = {https://ui.adsabs.harvard.edu/abs/2025NatAs...9.1476C}
}

@Inbook{Pinti2020,
author="Pinti, Daniele L.",
editor="Gargaud, Muriel
and Irvine, William M.
and Amils, Ricardo
and Claeys, Philippe
and Cleaves, Henderson James
and Gerin, Maryvonne
and Rouan, Daniel
and Spohn, Tilman
and Tirard, St{\'e}phane
and Viso, Michel",
title="Deuterium/Hydrogen Ratio (in Encyclopedia of Astrobiology)",
booktitle="Encyclopedia of Astrobiology",
year="2020",
publisher="Springer Berlin Heidelberg",
address="Berlin, Heidelberg",
pages="1--3",
isbn="978-3-642-27833-4",
doi="10.1007/978-3-642-27833-4_418-4",
url="https://doi.org/10.1007/978-3-642-27833-4_418-4"
}

@BOOK{Maeder2009,
       author = {{Maeder}, Andr{\'e}},
        title = "{Physics, Formation and Evolution of Rotating Stars}",
        publisher = "Springer Berlin Heidelberg",
         year = 2009,
          doi = {10.1007/978-3-540-76949-1},
       adsurl = {https://ui.adsabs.harvard.edu/abs/2009pfer.book.....M}
}

@INPROCEEDINGS{Smalley2004,
       author = {{Smalley}, Barry},
        title = "{Observations of convection in A-type stars}",
    booktitle = {The A-Star Puzzle},
         year = 2004,
       editor = {{Zverko}, Juraj and {Ziznovsky}, Jozef and {Adelman}, Saul J. and {Weiss}, Werner W.},
       series = {IAU Symposium},
       volume = {224},
        month = dec,
        pages = {131-138},
          doi = {10.1017/S1743921304004478},
archivePrefix = {arXiv},
       eprint = {astro-ph/0408222},
 primaryClass = {astro-ph},
       adsurl = {https://ui.adsabs.harvard.edu/abs/2004IAUS..224..131S}
}

@ARTICLE{Andrievsky2023,
       author = {{Andrievsky}, Sergei M. and {Kovtyukh}, Valery V.},
        title = "{Probing the Przybylski star for deuterium}",
      journal = {Astronomische Nachrichten},
         year = 2023,
        month = mar,
       volume = {344},
       number = {3},
          eid = {e20220133},
        pages = {e20220133},
          doi = {10.1002/asna.20220133},
archivePrefix = {arXiv},
       eprint = {2302.02487},
 primaryClass = {astro-ph.SR},
       adsurl = {https://ui.adsabs.harvard.edu/abs/2023AN....34420133A}
}

@ARTICLE{PeimbertI_1965,
       author = {{Peimbert}, Manuel and {Wallerstein}, George},
        title = "{A Search for Deuterium in Stellar Spectra. I. Magnetic Stars.}",
      journal = {\apj},
         year = 1965,
        month = feb,
       volume = {141},
        pages = {582},
          doi = {10.1086/148146},
       adsurl = {https://ui.adsabs.harvard.edu/abs/1965ApJ...141..582P}
}

@ARTICLE{PeimbertII_1965,
       author = {{Peimbert}, Manuel and {Wallerstein}, George},
        title = "{A Search for Deuterium in Stellar Spectra. II. Normal Stars of Types b, a, and F.}",
      journal = {\apj},
         year = 1965,
        month = oct,
       volume = {142},
        pages = {1024},
          doi = {10.1086/148371},
       adsurl = {https://ui.adsabs.harvard.edu/abs/1965ApJ...142.1024P}
}

@ARTICLE{Ferlet1983,
       author = {{Ferlet}, R. and {Dennefeld}, M. and {Spite}, M.},
        title = "{Stellar deuterium abundance : a new upper limit in Canopus.}",
      journal = {\aap},
         year = 1983,
        month = aug,
       volume = {124},
        pages = {172-174},
       adsurl = {https://ui.adsabs.harvard.edu/abs/1983A&A...124..172F}
}

@ARTICLE{Peimbert1981,
       author = {{Peimbert}, M. and {Wallerstein}, G. and {Pilachowski}, C.~A.},
        title = "{An upper limit for the deuterium abundance in Canopus}",
      journal = {\aap},
         year = 1981,
        month = dec,
       volume = {104},
       number = {1},
        pages = {72-74},
       adsurl = {https://ui.adsabs.harvard.edu/abs/1981A&A...104...72P}
}

@ARTICLE{Fairlambetal2015,
       author = {{Fairlamb}, J.~R. and {Oudmaijer}, R.~D. and {Mendigut{\'\i}a}, I. and {Ilee}, J.~D. and {van den Ancker}, M.~E.},
        title = "{A spectroscopic survey of Herbig Ae/Be stars with X-shooter - I. Stellar parameters and accretion rates}",
      journal = {\mnras},
         year = 2015,
        month = oct,
       volume = {453},
       number = {1},
        pages = {976-1001},
          doi = {10.1093/mnras/stv1576},
archivePrefix = {arXiv},
       eprint = {1507.05967},
 primaryClass = {astro-ph.SR},
       adsurl = {https://ui.adsabs.harvard.edu/abs/2015MNRAS.453..976F}
}

@ARTICLE{Kraletal2017,
       author = {{Kral}, Quentin and {Matr{\`a}}, Luca and {Wyatt}, Mark C. and {Kennedy}, Grant M.},
        title = "{Predictions for the secondary CO, C and O gas content of debris discs from the destruction of volatile-rich planetesimals}",
      journal = {\mnras},
         year = 2017,
        month = jul,
       volume = {469},
       number = {1},
        pages = {521-550},
          doi = {10.1093/mnras/stx730},
archivePrefix = {arXiv},
       eprint = {1703.10693},
 primaryClass = {astro-ph.EP},
       adsurl = {https://ui.adsabs.harvard.edu/abs/2017MNRAS.469..521K}
}

@ARTICLE{Kraletal2019,
       author = {{Kral}, Quentin and {Marino}, Sebastian and {Wyatt}, Mark C. and {Kama}, Mihkel and {Matr{\`a}}, Luca},
        title = "{Imaging [CI] around HD 131835: reinterpreting young debris discs with protoplanetary disc levels of CO gas as shielded secondary discs}",
      journal = {\mnras},
         year = 2019,
        month = nov,
       volume = {489},
       number = {4},
        pages = {3670-3691},
          doi = {10.1093/mnras/sty2923},
archivePrefix = {arXiv},
       eprint = {1811.08439},
 primaryClass = {astro-ph.EP},
       adsurl = {https://ui.adsabs.harvard.edu/abs/2019MNRAS.489.3670K}
}

@ARTICLE{JermynKama2018,
       author = {{Jermyn}, Adam S. and {Kama}, Mihkel},
        title = "{Stellar photospheric abundances as a probe of discs and planets}",
      journal = {\mnras},
         year = 2018,
        month = jun,
       volume = {476},
       number = {4},
        pages = {4418-4434},
          doi = {10.1093/mnras/sty429},
archivePrefix = {arXiv},
       eprint = {1804.06414},
 primaryClass = {astro-ph.SR},
       adsurl = {https://ui.adsabs.harvard.edu/abs/2018MNRAS.476.4418J}
}

@ARTICLE{Foreman-Mackey2013,
       author = {{Foreman-Mackey}, Daniel and {Hogg}, David W. and {Lang}, Dustin and {Goodman}, Jonathan},
        title = "{emcee: The MCMC Hammer}",
      journal = {\pasp},
         year = 2013,
        month = mar,
       volume = {125},
       number = {925},
        pages = {306},
          doi = {10.1086/670067},
archivePrefix = {arXiv},
       eprint = {1202.3665},
 primaryClass = {astro-ph.IM},
       adsurl = {https://ui.adsabs.harvard.edu/abs/2013PASP..125..306F}
}

@ARTICLE{Foreman-Mackey2016,
       author = {{Foreman-Mackey}, Daniel},
        title = "{corner.py: Scatterplot matrices in Python}",
      journal = {The Journal of Open Source Software},
         year = 2016,
        month = jun,
       volume = {1},
        pages = {24},
          doi = {10.21105/joss.00024},
       adsurl = {https://ui.adsabs.harvard.edu/abs/2016JOSS....1...24F}
}

@INPROCEEDINGS{Donati2003,
       author = {{Donati}, J.-F.},
        title = "{ESPaDOnS: An Echelle SpectroPolarimetric Device for the Observation of Stars at CFHT}",
    booktitle = {Solar Polarization},
         year = 2003,
       editor = {{Trujillo-Bueno}, Javier and {Sanchez Almeida}, Jorge},
       series = {Astronomical Society of the Pacific Conference Series},
       volume = {307},
        month = jan,
        pages = {41},
       adsurl = {https://ui.adsabs.harvard.edu/abs/2003ASPC..307...41D}
}

@ARTICLE{Fossati2011,
       author = {{Fossati}, L. and {Ryabchikova}, T. and {Shulyak}, D.~V. and {Haswell}, C.~A. and {Elmasli}, A. and {Pandey}, C.~P. and {Barnes}, T.~G. and {Zwintz}, K.},
        title = "{The accuracy of stellar atmospheric parameter determinations: a case study with HD 32115 and HD 37594}",
      journal = {\mnras},
         year = 2011,
        month = oct,
       volume = {417},
       number = {1},
        pages = {495-507},
          doi = {10.1111/j.1365-2966.2011.19289.x},
archivePrefix = {arXiv},
       eprint = {1106.4406},
 primaryClass = {astro-ph.SR},
       adsurl = {https://ui.adsabs.harvard.edu/abs/2011MNRAS.417..495F}
}

@ARTICLE{Fossati2009,
       author = {{Fossati}, L. and {Ryabchikova}, T. and {Bagnulo}, S. and {Alecian}, E. and {Grunhut}, J. and {Kochukhov}, O. and {Wade}, G.},
        title = "{The chemical abundance analysis of normal early A- and late B-type stars}",
      journal = {\aap},
         year = 2009,
        month = sep,
       volume = {503},
       number = {3},
        pages = {945-962},
          doi = {10.1051/0004-6361/200811561},
archivePrefix = {arXiv},
       eprint = {0906.5269},
 primaryClass = {astro-ph.SR},
       adsurl = {https://ui.adsabs.harvard.edu/abs/2009A&A...503..945F}
}

@ARTICLE{Piskunov1995,
       author = {{Piskunov}, N.~E. and {Kupka}, F. and {Ryabchikova}, T.~A. and {Weiss}, W.~W. and {Jeffery}, C.~S.},
        title = "{VALD: The Vienna Atomic Line Data Base.}",
      journal = {\aaps},
         year = 1995,
        month = sep,
       volume = {112},
        pages = {525},
       adsurl = {https://ui.adsabs.harvard.edu/abs/1995A&AS..112..525P},
         note = {\url{https://vald.astro.uu.se/}}
}

@ARTICLE{Ryabchikova1997,
       author = {{Ryabchikova}, T.~A. and {Piskunov}, N.~E. and {Kupka}, F. and {Weiss}, W.~W.},
        title = "{The Vienna Atomic Line Database : Present State and Future Development}",
      journal = {Baltic Astronomy},
         year = 1997,
        month = mar,
       volume = {6},
        pages = {244-247},
          doi = {10.1515/astro-1997-0216},
       adsurl = {https://ui.adsabs.harvard.edu/abs/1997BaltA...6..244R}
}

@ARTICLE{Kupka1999,
       author = {{Kupka}, F. and {Piskunov}, N. and {Ryabchikova}, T.~A. and {Stempels}, H.~C. and {Weiss}, W.~W.},
        title = "{VALD-2: Progress of the Vienna Atomic Line Data Base}",
      journal = {\aaps},
         year = 1999,
        month = jul,
       volume = {138},
        pages = {119-133},
          doi = {10.1051/aas:1999267},
       adsurl = {https://ui.adsabs.harvard.edu/abs/1999A&AS..138..119K}
}

@ARTICLE{Kupka2000,
       author = {{Kupka}, F.~G. and {Ryabchikova}, T.~A. and {Piskunov}, N.~E. and {Stempels}, H.~C. and {Weiss}, W.~W.},
        title = "{VALD-2 -- The New Vienna Atomic Line Database}",
      journal = {Baltic Astronomy},
         year = 2000,
        month = jan,
       volume = {9},
        pages = {590-594},
          doi = {10.1515/astro-2000-0420},
       adsurl = {https://ui.adsabs.harvard.edu/abs/2000BaltA...9..590K}
}

@ARTICLE{Ryabchikova2015,
       author = {{Ryabchikova}, T. and {Piskunov}, N. and {Kurucz}, R.~L. and {Stempels}, H.~C. and {Heiter}, U. and {Pakhomov}, Yu and {Barklem}, P.~S.},
        title = "{A major upgrade of the VALD database}",
      journal = {\physscr},
         year = 2015,
        month = may,
       volume = {90},
       number = {5},
          eid = {054005},
        pages = {054005},
          doi = {10.1088/0031-8949/90/5/054005},
       adsurl = {https://ui.adsabs.harvard.edu/abs/2015PhyS...90e4005R}
}

@ARTICLE{Pakhomov2019,
       author = {{Pakhomov}, Yu. V. and {Ryabchikova}, T.~A. and {Piskunov}, N.~E.},
        title = "{Hyperfine Splitting in the VALD Database of Spectral-line Parameters}",
      journal = {Astronomy Reports},
         year = 2019,
        month = dec,
       volume = {63},
       number = {12},
        pages = {1010-1021},
          doi = {10.1134/S1063772919120047},
archivePrefix = {arXiv},
       eprint = {1911.03189},
 primaryClass = {astro-ph.IM},
       adsurl = {https://ui.adsabs.harvard.edu/abs/2019ARep...63.1010P}
}

@ARTICLE{GAIADR3,
       author = {{Gaia Collaboration} and {Vallenari}, A. and {Brown}, A.~G.~A. and {Prusti}, T. and {de Bruijne}, J.~H.~J. and {Arenou}, F. and {Babusiaux}, C. and {Biermann}, M. and {Creevey}, O.~L. and {Ducourant}, C. and {Evans}, D.~W. and {Eyer}, L. and {Guerra}, R. and {Hutton}, A. and {Jordi}, C. and {Klioner}, S.~A. and {Lammers}, U.~L. and {Lindegren}, L. and {Luri}, X. and {Mignard}, F. and {Panem}, C. and {Pourbaix}, D. and {Randich}, S. and {Sartoretti}, P. and {Soubiran}, C. and {Tanga}, P. and {Walton}, N.~A. and {Bailer-Jones}, C.~A.~L. and {Bastian}, U. and {Drimmel}, R. and {Jansen}, F. and {Katz}, D. and {Lattanzi}, M.~G. and {van Leeuwen}, F. and {Bakker}, J. and {Cacciari}, C. and {Casta{\~n}eda}, J. and {De Angeli}, F. and {Fabricius}, C. and {Fouesneau}, M. and {Fr{\'e}mat}, Y. and {Galluccio}, L. and {Guerrier}, A. and {Heiter}, U. and {Masana}, E. and {Messineo}, R. and {Mowlavi}, N. and {Nicolas}, C. and {Nienartowicz}, K. and {Pailler}, F. and {Panuzzo}, P. and {Riclet}, F. and {Roux}, W. and {Seabroke}, G.~M. and {Sordo}, R. and {Th{\'e}venin}, F. and {Gracia-Abril}, G. and {Portell}, J. and {Teyssier}, D. and {Altmann}, M. and {Andrae}, R. and {Audard}, M. and {Bellas-Velidis}, I. and {Benson}, K. and {Berthier}, J. and {Blomme}, R. and {Burgess}, P.~W. and {Busonero}, D. and {Busso}, G. and {C{\'a}novas}, H. and {Carry}, B. and {Cellino}, A. and {Cheek}, N. and {Clementini}, G. and {Damerdji}, Y. and {Davidson}, M. and {de Teodoro}, P. and {Nu{\~n}ez Campos}, M. and {Delchambre}, L. and {Dell'Oro}, A. and {Esquej}, P. and {Fern{\'a}ndez-Hern{\'a}ndez}, J. and {Fraile}, E. and {Garabato}, D. and {Garc{\'\i}a-Lario}, P. and {Gosset}, E. and {Haigron}, R. and {Halbwachs}, J.-L. and {Hambly}, N.~C. and {Harrison}, D.~L. and {Hern{\'a}ndez}, J. and {Hestroffer}, D. and {Hodgkin}, S.~T. and {Holl}, B. and {Jan{\ss}en}, K. and {Jevardat de Fombelle}, G. and {Jordan}, S. and {Krone-Martins}, A. and {Lanzafame}, A.~C. and {L{\"o}ffler}, W. and {Marchal}, O. and {Marrese}, P.~M. and {Moitinho}, A. and {Muinonen}, K. and {Osborne}, P. and {Pancino}, E. and {Pauwels}, T. and {Recio-Blanco}, A. and {Reyl{\'e}}, C. and {Riello}, M. and {Rimoldini}, L. and {Roegiers}, T. and {Rybizki}, J. and {Sarro}, L.~M. and {Siopis}, C. and {Smith}, M. and {Sozzetti}, A. and {Utrilla}, E. and {van Leeuwen}, M. and {Abbas}, U. and {{\'A}brah{\'a}m}, P. and {Abreu Aramburu}, A. and {Aerts}, C. and {Aguado}, J.~J. and {Ajaj}, M. and {Aldea-Montero}, F. and {Altavilla}, G. and {{\'A}lvarez}, M.~A. and {Alves}, J. and {Anders}, F. and {Anderson}, R.~I. and {Anglada Varela}, E. and {Antoja}, T. and {Baines}, D. and {Baker}, S.~G. and {Balaguer-N{\'u}{\~n}ez}, L. and {Balbinot}, E. and {Balog}, Z. and {Barache}, C. and {Barbato}, D. and {Barros}, M. and {Barstow}, M.~A. and {Bartolom{\'e}}, S. and {Bassilana}, J.-L. and {Bauchet}, N. and {Becciani}, U. and {Bellazzini}, M. and {Berihuete}, A. and {Bernet}, M. and {Bertone}, S. and {Bianchi}, L. and {Binnenfeld}, A. and {Blanco-Cuaresma}, S. and {Blazere}, A. and {Boch}, T. and {Bombrun}, A. and {Bossini}, D. and {Bouquillon}, S. and {Bragaglia}, A. and {Bramante}, L. and {Breedt}, E. and {Bressan}, A. and {Brouillet}, N. and {Brugaletta}, E. and {Bucciarelli}, B. and {Burlacu}, A. and {Butkevich}, A.~G. and {Buzzi}, R. and {Caffau}, E. and {Cancelliere}, R. and {Cantat-Gaudin}, T. and {Carballo}, R. and {Carlucci}, T. and {Carnerero}, M.~I. and {Carrasco}, J.~M. and {Casamiquela}, L. and {Castellani}, M. and {Castro-Ginard}, A. and {Chaoul}, L. and {Charlot}, P. and {Chemin}, L. and {Chiaramida}, V. and {Chiavassa}, A. and {Chornay}, N. and {Comoretto}, G. and {Contursi}, G. and {Cooper}, W.~J. and {Cornez}, T. and {Cowell}, S. and {Crifo}, F. and {Cropper}, M. and {Crosta}, M. and {Crowley}, C. and {Dafonte}, C. and {Dapergolas}, A. and {David}, M. and {David}, P. and {de Laverny}, P. and {De Luise}, F. and {De March}, R.},
        title = "{Gaia Data Release 3. Summary of the content and survey properties}",
      journal = {\aap},
         year = 2023,
        month = jun,
       volume = {674},
          eid = {A1},
        pages = {A1},
          doi = {10.1051/0004-6361/202243940},
archivePrefix = {arXiv},
       eprint = {2208.00211},
 primaryClass = {astro-ph.GA},
       adsurl = {https://ui.adsabs.harvard.edu/abs/2023A&A...674A...1G}
}

@misc{NIST_ASD,
  author = {{Kramida, A.} and {Ralchenko, Yu.} and {Reader, J.} and {NIST ASD Team}},
  title = {{NIST Atomic Spectra Database (ver. 5.12)}},
  year = {2024},
  howpublished = {\url{https://physics.nist.gov/asd}},
  note = {National Institute of Standards and Technology, Gaithersburg, MD}
}

@ARTICLE{PolarBase,
       author = {{Petit}, P. and {Louge}, T. and {Th{\'e}ado}, S. and {Paletou}, F. and {Manset}, N. and {Morin}, J. and {Marsden}, S.~C. and {Jeffers}, S.~V.},
        title = "{PolarBase: A Database of High-Resolution Spectropolarimetric Stellar Observations}",
      journal = {\pasp},
         year = 2014,
        month = may,
       volume = {126},
       number = {939},
        pages = {469},
          doi = {10.1086/676976},
archivePrefix = {arXiv},
       eprint = {1401.1082},
 primaryClass = {astro-ph.SR},
       adsurl = {https://ui.adsabs.harvard.edu/abs/2014PASP..126..469P}
}

@misc{Kurucz1993,
  author = {{Kurucz}, Robert},
  title = "{ATLAS9 Stellar Atmosphere Programs and 2 km/s grid.}",
  year = {1993},
  howpublished = {Kurucz CD-ROM No. 13. Cambridge, Mass.: Smithsonian Astrophysical Observatory},
  note = {\url{http://kurucz.harvard.edu/programs/atlas9/}},
}

@misc{CastelliKurucz2004,
      title={New Grids of ATLAS9 Model Atmospheres}, 
      author={F. Castelli and R. L. Kurucz},
      year={2004},
      eprint={astro-ph/0405087},
      archivePrefix={arXiv},
      primaryClass={astro-ph}
}

@ARTICLE{Wade2001,
       author = {{Wade}, G.~A. and {Bagnulo}, S. and {Kochukhov}, O. and {Landstreet}, J.~D. and {Piskunov}, N. and {Stift}, M.~J.},
        title = "{LTE spectrum synthesis in magnetic stellar atmospheres. The interagreement of three independent polarised radiative transfer codes}",
      journal = {\aap},
         year = 2001,
        month = jul,
       volume = {374},
        pages = {265-279},
          doi = {10.1051/0004-6361:20010735},
       adsurl = {https://ui.adsabs.harvard.edu/abs/2001A&A...374..265W}
}

@ARTICLE{Landstreet1988,
       author = {{Landstreet}, J.~D.},
        title = "{The Magnetic Field and Abundance Distribution Geometry of the Peculiar A Star 53 Camelopardalis}",
      journal = {\apj},
         year = 1988,
        month = mar,
       volume = {326},
        pages = {967},
          doi = {10.1086/166155},
       adsurl = {https://ui.adsabs.harvard.edu/abs/1988ApJ...326..967L}
}

@ARTICLE{Folsom2012,
       author = {{Folsom}, C.~P. and {Bagnulo}, S. and {Wade}, G.~A. and {Alecian}, E. and {Landstreet}, J.~D. and {Marsden}, S.~C. and {Waite}, I.~A.},
        title = "{Chemical abundances of magnetic and non-magnetic Herbig Ae/Be stars}",
      journal = {\mnras},
         year = 2012,
        month = may,
       volume = {422},
       number = {3},
        pages = {2072-2101},
          doi = {10.1111/j.1365-2966.2012.20718.x},
archivePrefix = {arXiv},
       eprint = {1202.1845},
 primaryClass = {astro-ph.SR},
       adsurl = {https://ui.adsabs.harvard.edu/abs/2012MNRAS.422.2072F}
}

@ARTICLE{Folsom2022,
       author = {{Folsom}, Colin P. and {Kama}, Mihkel and {Eenm{\"a}e}, T{\~o}nis and {Kolka}, Indrek and {Aret}, Anna and {Checha}, Vitalii and {Kasikov}, Anni and {Leedj{\"a}rv}, Laurits and {Ramler}, Heleri},
        title = "{A rare phosphorus-rich star in an eclipsing binary from TESS}",
      journal = {\aap},
         year = 2022,
        month = feb,
       volume = {658},
          eid = {A105},
        pages = {A105},
          doi = {10.1051/0004-6361/202142124},
archivePrefix = {arXiv},
       eprint = {2111.07526},
 primaryClass = {astro-ph.SR},
       adsurl = {https://ui.adsabs.harvard.edu/abs/2022A&A...658A.105F}
}

@ARTICLE{Coutens2012,
       author = {{Coutens}, A. and {Vastel}, C. and {Caux}, E. and {Ceccarelli}, C. and {Bottinelli}, S. and {Wiesenfeld}, L. and {Faure}, A. and {Scribano}, Y. and {Kahane}, C.},
        title = "{A study of deuterated water in the low-mass protostar IRAS 16293-2422}",
      journal = {\aap},
         year = 2012,
        month = mar,
       volume = {539},
          eid = {A132},
        pages = {A132},
          doi = {10.1051/0004-6361/201117627},
archivePrefix = {arXiv},
       eprint = {1201.1785},
 primaryClass = {astro-ph.GA},
       adsurl = {https://ui.adsabs.harvard.edu/abs/2012A&A...539A.132C}
}

@ARTICLE{Tobin2023,
       author = {{Tobin}, John J. and {van't Hoff}, Merel L.~R. and {Leemker}, Margot and {van Dishoeck}, Ewine F. and {Paneque-Carre{\~n}o}, Teresa and {Furuya}, Kenji and {Harsono}, Daniel and {Persson}, Magnus V. and {Cleeves}, L. Ilsedore and {Sheehan}, Patrick D. and {Cieza}, Lucas},
        title = "{Deuterium-enriched water ties planet-forming disks to comets and protostars}",
      journal = {\nat},
         year = 2023,
        month = mar,
       volume = {615},
       number = {7951},
        pages = {227-230},
          doi = {10.1038/s41586-022-05676-z},
       adsurl = {https://ui.adsabs.harvard.edu/abs/2023Natur.615..227T}
}

@ARTICLE{Leemker2025,
       author = {{Leemker}, Margot and {Tobin}, John J. and {Facchini}, Stefano and {Curone}, Pietro and {Booth}, Alice S. and {Furuya}, Kenji and {van't Hoff}, Merel L.~R.},
        title = "{Pristine ices in a planet-forming disk revealed by heavy water}",
      journal = {Nature Astronomy},
         year = 2025,
        month = oct,
       volume = {9},
        pages = {1486-1494},
          doi = {10.1038/s41550-025-02663-y},
archivePrefix = {arXiv},
       eprint = {2510.19919},
 primaryClass = {astro-ph.EP},
       adsurl = {https://ui.adsabs.harvard.edu/abs/2025NatAs...9.1486L}
}

@ARTICLE{Slavicinska2025,
       author = {{Slavicinska}, Katerina and {Tychoniec}, {\L}ukasz and {Navarro}, Mar{\'\i}a Gabriela and {van Dishoeck}, Ewine F. and {Tobin}, John J. and {van Gelder}, Martijn L. and {Chen}, Yuan and {Boogert}, A.~C. Adwin and {Drechsler}, W. Blake and {Beuther}, Henrik and {Caratti o Garatti}, Alessio and {Megeath}, S. Thomas and {Klaassen}, Pamela and {Looney}, Leslie W. and {Kavanagh}, Patrick J. and {Brunken}, Nashanty G.~C. and {Sheehan}, Patrick and {Fischer}, William J.},
        title = "{HDO Ice Detected toward an Isolated Low-mass Protostar with JWST}",
      journal = {\apjl},
         year = 2025,
        month = jun,
       volume = {986},
       number = {2},
          eid = {L19},
        pages = {L19},
          doi = {10.3847/2041-8213/addb45},
archivePrefix = {arXiv},
       eprint = {2505.14686},
 primaryClass = {astro-ph.SR},
       adsurl = {https://ui.adsabs.harvard.edu/abs/2025ApJ...986L..19S}
}

@ARTICLE{Lane2026,
       author = {{Lane}, Kaitlyn T. and {Stephan}, Alexander P. and {Soares-Furtado}, Melinda and {Stassun}, Keivan G. and {Yarza}, Ricardo},
        title = "{Observable Metal Pollution in Main-sequence Stars: Simulations of Rocky Planets Engulfed by Stars in the 0.5 to 1.4 M$_{{\ensuremath{\odot}}}$ Range}",
      journal = {\apj},
         year = 2026,
        month = may,
       volume = {1003},
       number = {1},
          eid = {67},
        pages = {67},
          doi = {10.3847/1538-4357/ae5b9a},
archivePrefix = {arXiv},
       eprint = {2601.00949},
 primaryClass = {astro-ph.EP},
       adsurl = {https://ui.adsabs.harvard.edu/abs/2026ApJ..1003...67L}
}

@ARTICLE{Lellouch2001,
       author = {{Lellouch}, E. and {B{\'e}zard}, B. and {Fouchet}, T. and {Feuchtgruber}, H. and {Encrenaz}, T. and {de Graauw}, T.},
        title = "{The deuterium abundance in Jupiter and Saturn from ISO-SWS observations}",
      journal = {\aap},
         year = 2001,
        month = may,
       volume = {370},
        pages = {610-622},
          doi = {10.1051/0004-6361:20010259},
       adsurl = {https://ui.adsabs.harvard.edu/abs/2001A&A...370..610L}
}

@ARTICLE{Pierel2017,
       author = {{Pierel}, J.~D.~R. and {Nixon}, C.~A. and {Lellouch}, E. and {Fletcher}, L.~N. and {Bjoraker}, G.~L. and {Achterberg}, R.~K. and {B{\'e}zard}, B. and {Hesman}, B.~E. and {Irwin}, P.~G.~J. and {Flasar}, F.~M.},
        title = "{D/H Ratios on Saturn and Jupiter from Cassini CIRS}",
      journal = {\aj},
         year = 2017,
        month = nov,
       volume = {154},
       number = {5},
          eid = {178},
        pages = {178},
          doi = {10.3847/1538-3881/aa899d},
       adsurl = {https://ui.adsabs.harvard.edu/abs/2017AJ....154..178P}
}

@ARTICLE{Feuchtgruber1999,
       author = {{Feuchtgruber}, H. and {Lellouch}, E. and {B{\'e}zard}, B. and {Encrenaz}, Th. and {de Graauw}, Th. and {Davis}, G.~R.},
        title = "{Detection of HD in the atmospheres of Uranus and Neptune: a new determination of the D/H ratio}",
      journal = {\aap},
         year = 1999,
        month = jan,
       volume = {341},
        pages = {L17-L21},
       adsurl = {https://ui.adsabs.harvard.edu/abs/1999A&A...341L..17F}
}

@BOOK{Gray2005-Photospheres,
   author = {{Gray}, D.~F.},
    title = "{The Observation and Analysis of Stellar Photospheres}",
edition = {3rd},
publisher = {Cambridge University Press},
address = {Cambridge, UK},
     year = 2005,
    month = sep,
}

@BOOK{Griffiths1994-quantum-intro,
   author = {{Griffiths}, D.~J.},
    title = "{Introduction to quantum mechanics}",
publisher = {Prentics Hall},
address = {Upper Saddle River, New Jersey},
     year = 1994,
}

@ARTICLE{Irwin1981-partition-functions,
       author = {{Irwin}, A.~W.},
        title = "{Polynomial partition function approximations of 344 atomic and molecular species.}",
      journal = {\apjs},
         year = 1981,
        month = apr,
       volume = {45},
        pages = {621-633},
          doi = {10.1086/190730},
       adsurl = {https://ui.adsabs.harvard.edu/abs/1981ApJS...45..621I}
}

@ARTICLE{Lemke1997,
       author = {{Lemke}, M.},
        title = "{Extended VCS Stark broadening tables for hydrogen -- Lyman to Brackett series}",
      journal = {\aaps},
         year = 1997,
        month = apr,
       volume = {122},
        pages = {285-292},
          doi = {10.1051/aas:1997134},
       adsurl = {https://ui.adsabs.harvard.edu/abs/1997A&AS..122..285L}
}

@ARTICLE{Vidal1973,
       author = {{Vidal}, C.~R. and {Cooper}, J. and {Smith}, E.~W.},
        title = "{Hydrogen Stark-Broadening Tables}",
      journal = {\apjs},
         year = 1973,
        month = jan,
       volume = {25},
        pages = {37},
          doi = {10.1086/190264},
       adsurl = {https://ui.adsabs.harvard.edu/abs/1973ApJS...25...37V}
}

@ARTICLE{AliGriem1965,
       author = {{Ali}, A.~W. and {Griem}, H.~R.},
        title = "{Theory of Resonance Broadening of Spectral Lines by Atom-Atom Impacts}",
      journal = {Physical Review},
         year = 1965,
        month = nov,
       volume = {140},
       number = {4A},
        pages = {1044-1049},
          doi = {10.1103/PhysRev.140.A1044},
       adsurl = {https://ui.adsabs.harvard.edu/abs/1965PhRv..140.1044A}
}

@ARTICLE{AliGriem1966,
       author = {{Ali}, A.~W. and {Griem}, H.~R.},
        title = "{Theory of Resonance Broadening of Spectral Lines by Atom-Atom Impacts}",
      journal = {Physical Review},
         year = 1966,
        month = apr,
       volume = {144},
       number = {1},
        pages = {366-366},
          doi = {10.1103/PhysRev.144.366},
       adsurl = {https://ui.adsabs.harvard.edu/abs/1966PhRv..144..366A}
}

@ARTICLE{Mendigutiaetal2012,
       author = {{Mendigut{\'\i}a}, I. and {Mora}, A. and {Montesinos}, B. and {Eiroa}, C. and {Meeus}, G. and {Mer{\'\i}n}, B. and {Oudmaijer}, R.~D.},
        title = "{Accretion-related properties of Herbig Ae/Be stars. Comparison with T Tauris}",
      journal = {\aap},
         year = 2012,
        month = jul,
       volume = {543},
          eid = {A59},
        pages = {A59},
          doi = {10.1051/0004-6361/201219110},
archivePrefix = {arXiv},
       eprint = {1205.4734},
 primaryClass = {astro-ph.SR},
       adsurl = {https://ui.adsabs.harvard.edu/abs/2012A&A...543A..59M}
}

@ARTICLE{Kama2015,
       author = {{Kama}, M. and {Folsom}, C.~P. and {Pinilla}, P.},
        title = "{Fingerprints of giant planets in the photospheres of Herbig stars}",
      journal = {\aap},
         year = 2015,
        month = oct,
       volume = {582},
          eid = {L10},
        pages = {L10},
          doi = {10.1051/0004-6361/201527094},
archivePrefix = {arXiv},
       eprint = {1509.02741},
 primaryClass = {astro-ph.SR},
       adsurl = {https://ui.adsabs.harvard.edu/abs/2015A&A...582L..10K}
}

@ARTICLE{Kama2016,
       author = {{Kama}, M. and {Bruderer}, S. and {van Dishoeck}, E.~F. and {Hogerheijde}, M. and {Folsom}, C.~P. and {Miotello}, A. and {Fedele}, D. and {Belloche}, A. and {G{\"u}sten}, R. and {Wyrowski}, F.},
        title = "{Volatile-carbon locking and release in protoplanetary disks. A study of TW Hya and HD 100546}",
      journal = {\aap},
         year = 2016,
        month = aug,
       volume = {592},
          eid = {A83},
        pages = {A83},
          doi = {10.1051/0004-6361/201526991},
archivePrefix = {arXiv},
       eprint = {1605.05093},
 primaryClass = {astro-ph.EP},
       adsurl = {https://ui.adsabs.harvard.edu/abs/2016A&A...592A..83K}
}

@ARTICLE{Borthakur2025,
       author = {{Borthakur}, Sandipan P.~D. and {Kama}, Mihkel and {Fossati}, Luca and {Kral}, Quentin and {Folsom}, Colin P. and {Teske}, Johanna and {Aret}, Anna},
        title = "{Abundance analysis of stars hosting gas-rich debris discs}",
      journal = {\aap},
         year = 2025,
        month = may,
       volume = {697},
          eid = {A59},
        pages = {A59},
          doi = {10.1051/0004-6361/202452840},
archivePrefix = {arXiv},
       eprint = {2503.03614},
 primaryClass = {astro-ph.SR},
       adsurl = {https://ui.adsabs.harvard.edu/abs/2025A&A...697A..59B}
}

@ARTICLE{GuzmanDiaz2023,
       author = {{Guzm{\'a}n-D{\'\i}az}, J. and {Montesinos}, B. and {Mendigut{\'\i}a}, I. and {Kama}, M. and {Meeus}, G. and {Vioque}, M. and {Oudmaijer}, R.~D. and {Villaver}, E.},
        title = "{Relation between metallicities and spectral energy distributions of Herbig Ae/Be stars. A potential link with planet formation}",
      journal = {\aap},
         year = 2023,
        month = mar,
       volume = {671},
          eid = {A140},
        pages = {A140},
          doi = {10.1051/0004-6361/202245427},
archivePrefix = {arXiv},
       eprint = {2212.14022},
 primaryClass = {astro-ph.SR},
       adsurl = {https://ui.adsabs.harvard.edu/abs/2023A&A...671A.140G}
}

@ARTICLE{Murphy2021,
       author = {{Murphy}, Simon J. and {Joyce}, Meridith and {Bedding}, Timothy R. and {White}, Timothy R. and {Kama}, Mihkel},
        title = "{A precise asteroseismic age and metallicity for HD 139614: a pre-main-sequence star with a protoplanetary disc in Upper Centaurus-Lupus}",
      journal = {\mnras},
         year = 2021,
        month = apr,
       volume = {502},
       number = {2},
        pages = {1633-1646},
          doi = {10.1093/mnras/stab144},
archivePrefix = {arXiv},
       eprint = {2011.11821},
 primaryClass = {astro-ph.SR},
       adsurl = {https://ui.adsabs.harvard.edu/abs/2021MNRAS.502.1633M}
}

@ARTICLE{Krenn2021,
       author = {{Krenn}, A.~F. and {Fossati}, L. and {Kubyshkina}, D. and {Lammer}, H.},
        title = "{A critical assessment of the applicability of the energy-limited approximation for estimating exoplanetary mass-loss rates}",
      journal = {\aap},
         year = 2021,
        month = jun,
       volume = {650},
          eid = {A94},
        pages = {A94},
          doi = {10.1051/0004-6361/202140437},
archivePrefix = {arXiv},
       eprint = {2105.05858},
 primaryClass = {astro-ph.EP},
       adsurl = {https://ui.adsabs.harvard.edu/abs/2021A&A...650A..94K}
}

@ARTICLE{Owen2019,
       author = {{Owen}, James E.},
        title = "{Atmospheric Escape and the Evolution of Close-In Exoplanets}",
      journal = {Annual Review of Earth and Planetary Sciences},
         year = 2019,
        month = may,
       volume = {47},
        pages = {67-90},
          doi = {10.1146/annurev-earth-053018-060246},
archivePrefix = {arXiv},
       eprint = {1807.07609},
 primaryClass = {astro-ph.EP},
       adsurl = {https://ui.adsabs.harvard.edu/abs/2019AREPS..47...67O}
}

@ARTICLE{Wyttenbach2020,
       author = {{Wyttenbach}, A. and {Molli{\`e}re}, P. and {Ehrenreich}, D. and {Cegla}, H.~M. and {Bourrier}, V. and {Lovis}, C. and {Pino}, L. and {Allart}, R. and {Seidel}, J.~V. and {Hoeijmakers}, H.~J. and {Nielsen}, L.~D. and {Lavie}, B. and {Pepe}, F. and {Bonfils}, X. and {Snellen}, I.~A.~G.},
        title = "{Mass-loss rate and local thermodynamic state of the KELT-9 b thermosphere from the hydrogen Balmer series}",
      journal = {\aap},
         year = 2020,
        month = jun,
       volume = {638},
          eid = {A87},
        pages = {A87},
          doi = {10.1051/0004-6361/201937316},
archivePrefix = {arXiv},
       eprint = {2004.13733},
 primaryClass = {astro-ph.EP},
       adsurl = {https://ui.adsabs.harvard.edu/abs/2020A&A...638A..87W}
}

@ARTICLE{Anfinogentov2021-SoBAT-MCMC,
       author = {{Anfinogentov}, Sergey A. and {Nakariakov}, Valery M. and {Pascoe}, David J. and {Goddard}, Christopher R.},
        title = "{Solar Bayesian Analysis Toolkit{\textemdash}A New Markov Chain Monte Carlo IDL Code for Bayesian Parameter Inference}",
      journal = {\apjs},
         year = 2021,
        month = jan,
       volume = {252},
       number = {1},
          eid = {11},
        pages = {11},
          doi = {10.3847/1538-4365/abc5c1},
archivePrefix = {arXiv},
       eprint = {2005.05365},
 primaryClass = {astro-ph.SR},
       adsurl = {https://ui.adsabs.harvard.edu/abs/2021ApJS..252...11A}
}

@ARTICLE{Hogg2010-fitting-data,
       author = {{Hogg}, David W. and {Bovy}, Jo and {Lang}, Dustin},
        title = "{Data analysis recipes: Fitting a model to data}",
      journal = {arXiv e-prints},
         year = 2010,
        month = aug,
          eid = {arXiv:1008.4686},
        pages = {arXiv:1008.4686},
          doi = {10.48550/arXiv.1008.4686},
archivePrefix = {arXiv},
       eprint = {1008.4686},
 primaryClass = {astro-ph.IM},
       adsurl = {https://ui.adsabs.harvard.edu/abs/2010arXiv1008.4686H}
}

@ARTICLE{2022OAP....35...13A,
       author = {{Andrievsky}, S.~M.},
        title = "{An Enigma of the Przybylski Star}",
      journal = {Odessa Astronomical Publications},
         year = 2022,
        month = jan,
       volume = {35},
        pages = {13},
          doi = {10.18524/1810-4215.2022.35.268673},
       adsurl = {https://ui.adsabs.harvard.edu/abs/2022OAP....35...13A}
}

\onecolumn

\begin{appendix}

\section{Posterior probability distributions}\label{cornerplots}

\vspace{0pt}

\begin{figure}[!ht]
  \centering
  \includegraphics[width=\linewidth]{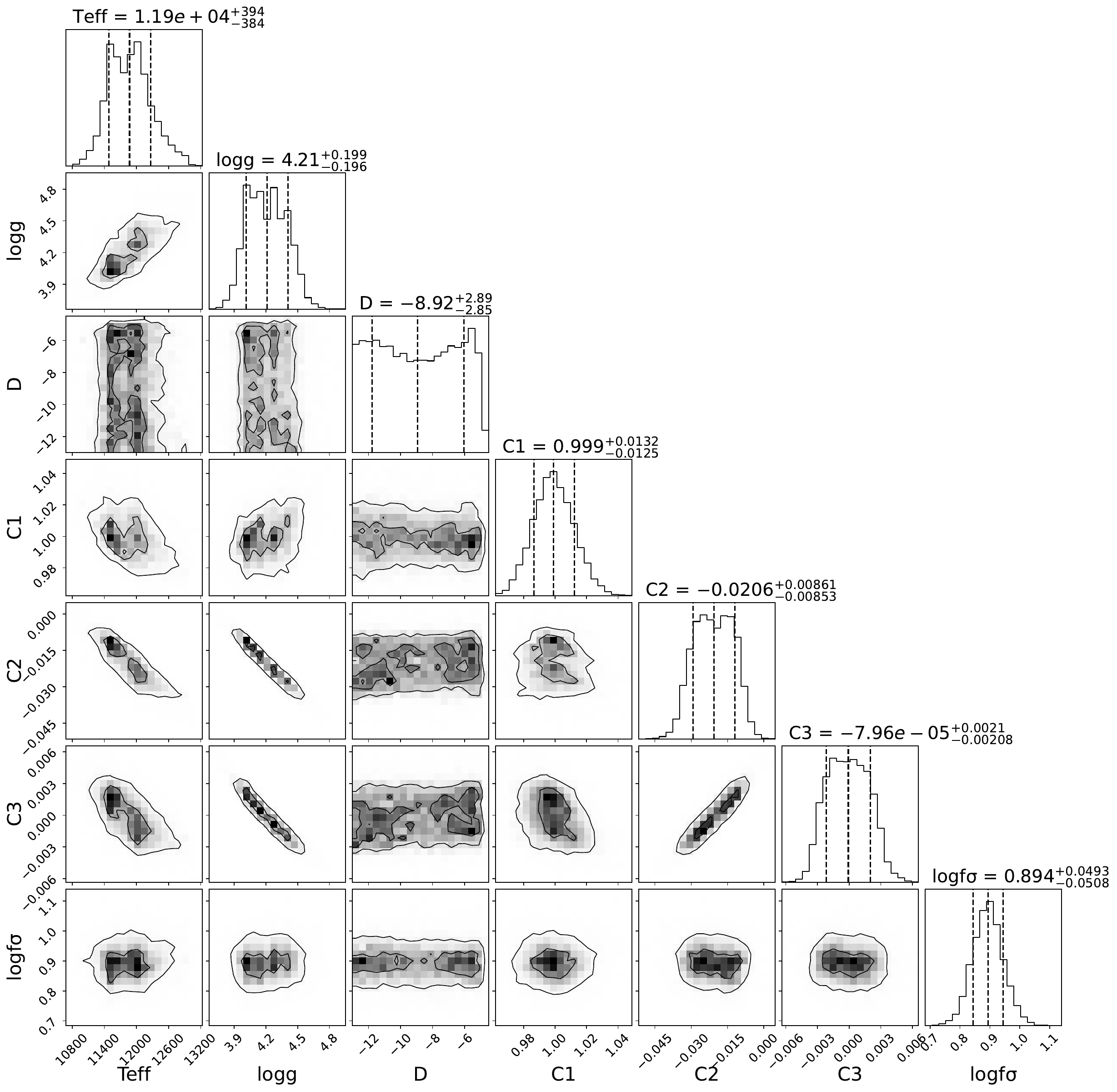}
  \caption{Posterior probability distributions obtained from H$\alpha$ profile fitting for 21\,Peg (see Sect.\ 3.2). One- and two-dimensional marginalised posteriors are shown, with contours enclosing 39.3\% and 86.5\% of the probability density and vertical lines marking the 16th, 50th, and 84th percentiles. }
  \label{fig:21PegHaCorner}
\end{figure}

\begin{figure}[!t]
  \centering
  \includegraphics[width=\linewidth]{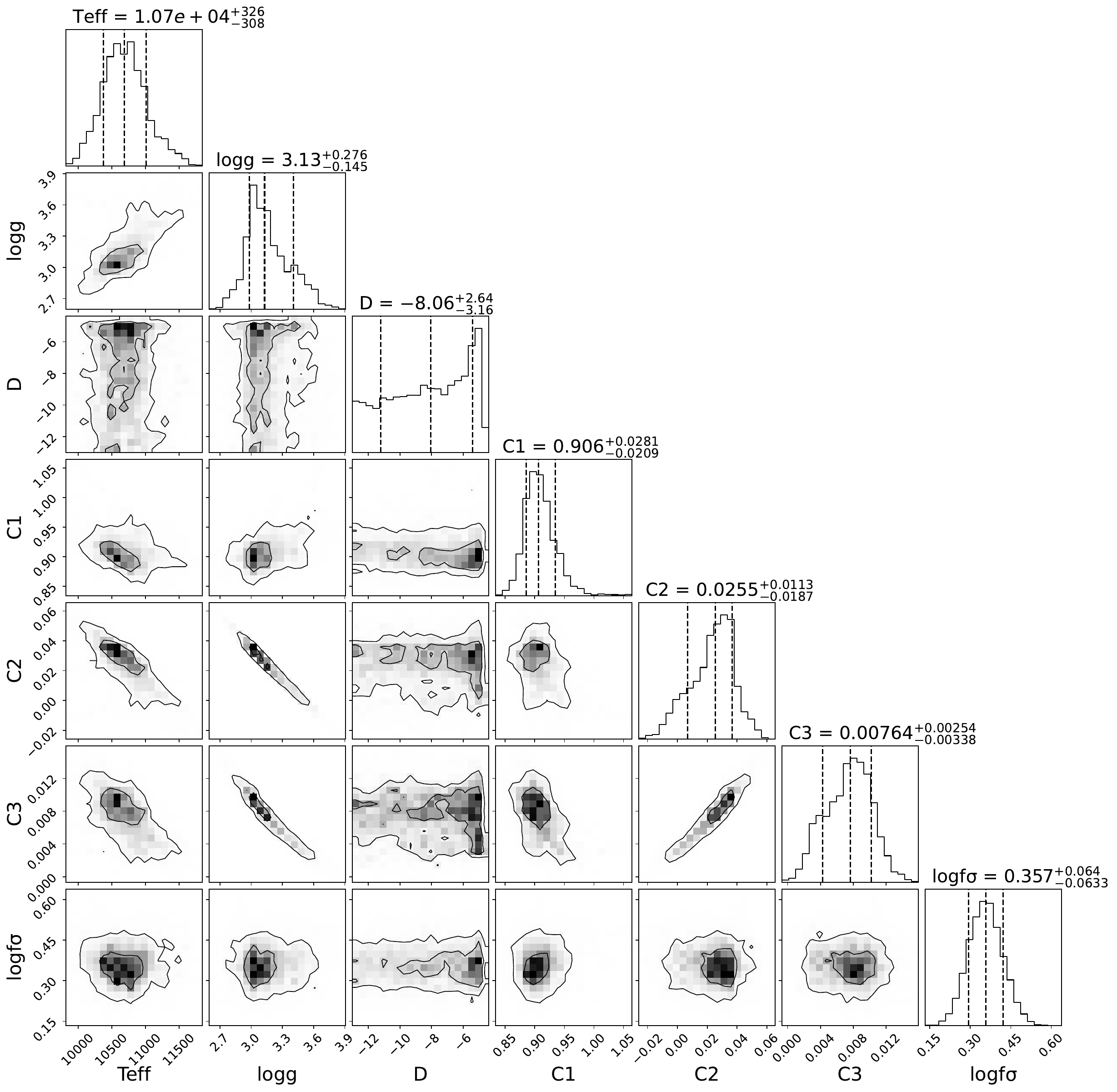}
  \caption{Posterior probability distributions obtained from H$\beta$ profile fitting for 21\,Peg, as in Fig.\,\ref{fig:21PegHaCorner}. }
  \label{fig:21PegHbCorner}
\end{figure}

\begin{figure}[!t]
  \centering
  \includegraphics[width=\linewidth]{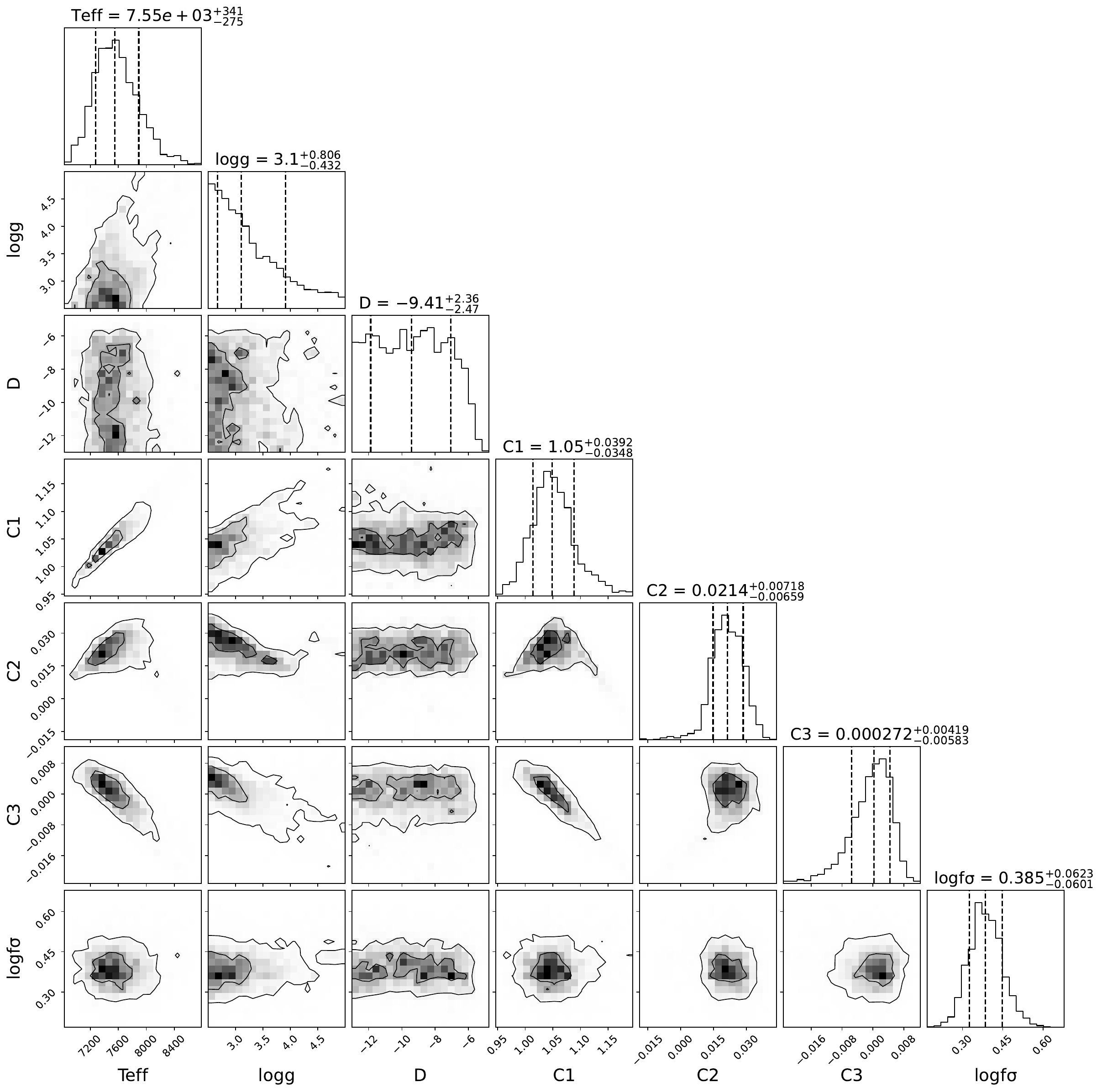}
  \caption{Posterior probability distributions obtained from H$\alpha$ profile fitting for HD\,32115 (see Sect.\, 3.2), as in Fig.\,\ref{fig:21PegHaCorner}.}
  \label{fig:HD32115Ha}
\end{figure}
\end{appendix}
\end{document}